\makeatletter\providecommand\declare@file@substitution[2]{}
\declare@file@substitution{revtex4-1.cls}{revtex4-2.cls}
\let\old@LaTeX\LaTeX
\documentclass[twocolumn,twocolappendix,tighten]{aastex631}
\ifpdf\pdfoutput=1\fi
\let\LaTeX\old@LaTeX
\makeatother
\renewcommand{\today}{2025 March 26}
\usepackage{amsmath,xspace,tabularx}
\usepackage[all]{hypcap}
\usepackage{xparse}

\usepackage[angle-symbol-over-decimal,unit-mode=text,
table-alignment-mode=format,table-number-alignment=center,
uncertainty-mode=separate,print-unity-mantissa=false,
separate-uncertainty-units=single,
range-units=single,range-phrase={\text{--}},list-units=single]{siunitx}

\hypersetup{bookmarksnumbered=true,bookmarksopen=true,
  pdfauthor=Xuchen Lin
}

\IfFileExists{latexml.sty}
  {\usepackage{latexml}}
  {\newif\iflatexml\latexmlfalse}
\ProvideDocumentCommand\published{m}{}
\ifxetex%
  \usepackage{xeCJK}
  \usepackage{unicode-math}
    [range="002F]
    [range={cal/Latin,bfcal/Latin},StylisticSet=1]
    [range={cal/latin,bfcal/latin}]
\else
  \usepackage[T1]{fontenc}
  \iflatexml\else
    \IfFileExists{newtx.sty}{%
      \usepackage{newtx}
    }{%
      \usepackage{newtxtext,newtxmath}
    }
    \usepackage{CJK}
  \fi
  \AtBeginDocument{\DeclareSIUnit[quantity-product={}]\degree{\text{\textdegree}}}
\fi
\AtBeginDocument{\DeclareSIUnit[quantity-product={},list-units=repeat]\percent{\text{\%}}}
\ProvideDocumentEnvironment{CJK*}{m m}{}{}
\makeatletter
\AtBeginDocument{
  \newcommand*{\cjkname}[3][gbsn]{#2 (\begin{CJK*}{UTF8}{#1}#3\ignorespacesafterend\end{CJK*})}
  \@ifpackageloaded{CJK}{}{
    \@ifpackageloaded{xeCJK}{}{
      \iflatexml\else
        \renewcommand*{\cjkname}[3][]{#2}
      \fi
    }
  }
}
\usepackage{etoolbox}
\newcount\replace@count
\newcommand*{\@checkappendixparam}[1]{%
  \def\@checkappendixparamtmp{#1}%
  \ifx\Hy@param\@checkappendixparamtmp
    \let\Hy@param\Hy@appendixstring
  \fi
}
\apptocmd{\appendix}{%
  \patchcmd{\hyper@makecurrent}{%
    \edef\Hy@param{#1}%
  }{%
    \edef\Hy@param{#1}%
    \@checkappendixparam{subsection}%
    \@checkappendixparam{subsubsection}%
  }{}{}%
}{}{}
\patchcmd{\HyOrg@appendix}{\onecolumngrid}{}{}{}
\hypersetup{
  linkcolor=blue,citecolor=blue,
  filecolor=blue,urlcolor=blue
}
\pretocmd{\maketitle}{
  \hypersetup{urlcolor=black}
}{}{}
\apptocmd{\maketitle}{
  \hypersetup{urlcolor=blue}
}{}{}
\patchcmd{\NAT@citex}
{\@citea\NAT@hyper@{%
    \NAT@nmfmt{\NAT@nm}%
    \hyper@natlinkbreak{\NAT@aysep\NAT@spacechar}{\@citeb\@extra@b@citeb}%
    \NAT@date}}
{\@citea\NAT@nmfmt{\NAT@nm}%
  \NAT@aysep\NAT@spacechar\NAT@hyper@{\NAT@date}}{}{}
\patchcmd{\NAT@citex}
{\@citea\NAT@hyper@{%
    \NAT@nmfmt{\NAT@nm}%
    \hyper@natlinkbreak{\NAT@spacechar\NAT@@open\if*#1*\else#1\NAT@spacechar\fi}%
    {\@citeb\@extra@b@citeb}%
    \NAT@date}}
{\@citea\NAT@nmfmt{\NAT@nm}%
  \NAT@spacechar\NAT@@open\if*#1*\else#1\NAT@spacechar\fi\NAT@hyper@{\NAT@date}}
{}{}
\iflatexml\providecommand{\onecolumngrid}{}\fi
\AtBeginEnvironment{thebibliography}{%
  \phantomsection%
  \addcontentsline{toc}{section}{\refname}%
}
\patchcmd{\thebibliography}{%
  \onecolumngrid}{}{}{}
\patchcmd{\thebibliography}{%
  \twocolumngrid}{}{}{}
\patchcmd{\thebibliography}{%
  \textwidth}{\linewidth}{}{}
\patchcmd{\thebibliography}{%
  \raggedright\small}{%
  \footnotesize%
}{}{}
\patchcmd{\thebibliography}{%
  \baselineskip=13pt plus 1pt}{%
  \baselineskip=9pt plus 1pt%
}{}{}
\patchcmd{\thebibliography}{%
  \baselineskip=13pt plus 1pt}{%
  \baselineskip=9pt plus 1pt%
}{}{}
  \patchcmd{\@makecaption}{\small}{\footnotesize}{}{}
  \advance\replace@count-1
\ifnum\replace@count>0\repeat
\patchcmd{\@facility}{%
  \large}{}{}{}
\patchcmd{\@facilities}{%
  \large}{}{}{}
\patchcmd{\@software}{%
  \large}{}{}{}
\patchcmd{\sec@upcase}{%
  \uppercase}{%
  \MakeTextUppercase%
}{}{}
\renewcommand\ltx@foottext[2]{%
  \begingroup
  \expandafter\ltx@make@current@footnote\expandafter{\@mpfn}{#1}%
  \@footnotetext{#2}
  \endgroup
}%
\renewcommand\@makefntext[1]{%
  \parindent 1em%
  \baselineskip 8.5\p@%
  \noindent
  \ifnum\c@footnote<100%
    \hb@xt@1.25em{\@makefnmark\hss}%
  \else%
    \hbox{\@makefnmark\ }%
  \fi#1%
}
\patchcmd{\acknowledgments}{%
  \begin{internallinenumbers}}{%
    \ifnumlines\begin{internallinenumbers}\fi%
      }{}{}
      \patchcmd{\endacknowledgments}{%
    \end{internallinenumbers}}{%
    \ifnumlines\end{internallinenumbers}\fi%
}{}{}
\patchcmd{\enddeluxetable}{%
  \null
}{}{}{}
\renewcommand\ApjSectionpenalty{\@M}
\renewcommand\aapr{\ref@jnl{A\&ARv}}
\renewcommand\nat{\ref@jnl{Natur}}
\makeatother

\newcommand*{\Mvir}{\ensuremath{M\tsb{vir}=\qty[parse-numbers=false]{10^{12.37}}{\Msun}}\xspace}

\newcommand{\numkcomb}{\num[parse-numbers=false]{0.548^{+0.078}_{-0.037}}\xspace}
\newcommand{\numkfaint}{\num[parse-numbers=false]{0.585^{+0.090}_{-0.019}}\xspace}
\newcommand{\numvcent}{\num{460.0}\xspace}
\newcommand{\qtyvcent}{\qty[parse-numbers=false]{\numvcent}{\km\per\s}\xspace}
\newcommand{\qtychanwidth}{\qty{5.153}{\km\per\s}\xspace}
\newcommand{\qtyNHilim}[1][\sim]{\qty{#1 3.2e18}{\per\cm\squared}\xspace}
\newcommand{\VLAcleanbeam}{\ensuremath{\ang{;;17.84}\times\ang{;;15.40}}\xspace}
\newcommand{\qtyVLAfwhm}[1][\sim]{\ang{;;#1 18}\xspace}
\newcommand{\qtyphysVLAfwhm}[1][\sim]{\qty{#1 0.7}{\kpc}\xspace}
\newcommand{\qtyFWHMlim}[1][\sim]{\ang{;;#1 39}\xspace}
\newcommand{\qtyphysFWHMlim}[1][\sim]{\qty{#1 1.5}{\kpc}\xspace}
\newcommand{\qtyFWHMlimup}[1][\sim]{\ang{;;#1 74}\xspace}
\newcommand{\qtyphysFWHMlimrange}[1][\sim]{\qtyrange{#1 1.5}{2.9}{\kpc}\xspace}
\newcommand{\qtyFWHMavg}[1][\sim]{\ang{;;#1 57}\xspace}
\newcommand{\qtyphysFWHMavg}[1][\sim]{\qty{2.2}{\kpc}\xspace}
\newcommand{\qtydiffusedivision}{\qty[parse-numbers=false]{10^{19.7}}{\per\cm\squared}\xspace}
\newcommand{\rnghighlydisperse}{\ensuremath{\qty[parse-numbers=false]{10^{1.9}}{\km\per\s}\approx\qty{80}{\km\per\s}}\xspace}
\newcommand{\qtysfr}{\qty[parse-numbers=false]{4.5^{+2.6}_{-1.6}}{\Msun\per\yr}}
\newcommand{\qtyRzs}{\qty{11.6}{\kpc}}
\newcommand{\qtycscgm}{\qty{156}{\km\per\s}}

\makeatletter
\renewcommand\ion[2]{\texorpdfstring{\text{#1\,\textsc{\@roman{#2}}}}{#1 \@roman{#2}}}
\makeatother
\newcommand*{\Hi}{\ion{H}{1}\xspace}
\newcommand*{\sbHi}[1][]{\tsb{\Hi\if\relax\detokenize{#1}\relax\else,#1\fi}\xspace}
\newcommand*{\tsb}[1]{\ensuremath{_{\text{#1}}}}

\newcommand*{\los}{LOS\xspace}
\newcommand*{\PA}{\text{P.A.}\xspace}
\newcommand*{\mprot}{\ensuremath{m\tsb{p}}\xspace}
\newcommand*{\supparen}[1]{\if\relax\detokenize{#1}\relax\else^{(#1)}\fi}
\newcommand*{\thres}[2][]{\ensuremath{\theta\tsb{#2}\supparen{#1}}}
\newcommand*{\goodness}[1][]{\ensuremath{\varrho\if\relax\detokenize{#1}\relax\else\tsb{#1}\fi}\xspace}
\newcommand*{\defim}[1][]{\ensuremath{\hat{m}_j\supparen{#1}}\xspace}
\newcommand*{\modim}[1][]{\ensuremath{m_j\supparen{#1}}\xspace}
\newcommand*{\resim}[2][]{\ensuremath{r_{\if\relax\detokenize{#2}\relax\else\text{#2},\fi j}\supparen{#1}}}
\newcommand*{\intim}{\ensuremath{d^\text{int}_j}\xspace}
\newcommand*{\intb}{\ensuremath{b^\text{int}}\xspace}
\newcommand*{\intres}[1][]{\resim[#1]{int}\xspace}
\newcommand*{\cleanb}{\ensuremath{b^\text{clean}}\xspace}
\newcommand*{\sdim}{\ensuremath{d^\text{sd}_j}\xspace}
\newcommand*{\sdb}{\ensuremath{b^\text{sd}}\xspace}
\newcommand*{\sdres}[1][]{\resim[#1]{sd}\xspace}
\newcommand*{\sigint}{\ensuremath{\sigma\tsb{int}}\xspace}
\newcommand*{\sigintout}[1][]{\ensuremath{\sigma\tsb{int,out}\supparen{#1}}\xspace}
\newcommand*{\sigsd}{\ensuremath{\sigma\tsb{sd}}\xspace}
\newcommand*{\Iv}{\ensuremath{I_v}\xspace}
\newcommand*{\Imax}{\ensuremath{I\tsb{max}}\xspace}
\newcommand*{\dMcool}{\ensuremath{\dot{M}\tsb{cool}}\xspace}
\newcommand*{\dMin}{\ensuremath{\dot{M}\tsb{in}}\xspace}
\newcommand*{\dMinTML}{\ensuremath{\dot{M}\tsb{in,TML}}\xspace}
\newcommand*{\bardMinTML}{\ensuremath{\overline{\dot{M}}\tsb{in,TML}}\xspace}
\newcommand*{\rring}{\ensuremath{r\tsb{ring}}\xspace}
\newcommand*{\Rzs}{\ensuremath{R_{25}}\xspace}
\newcommand*{\Rsoo}{\ensuremath{R_{500}}\xspace}
\newcommand*{\RHi}[1][]{\ensuremath{R\sbHi[#1]}\xspace}
\newcommand*{\vrot}{\ensuremath{v\tsb{rot}}\xspace}
\newcommand*{\vrad}{\ensuremath{v\tsb{rad}}\xspace}
\newcommand*{\vsys}{\ensuremath{v\tsb{sys}}\xspace}
\newcommand*{\vtot}{\ensuremath{v\tsb{tot}}\xspace}
\newcommand*{\vs}{\ensuremath{v\tsb{s}}\xspace}
\newcommand*{\vent}{\ensuremath{v\tsb{ent}}\xspace}

\newcommand*{\Sigdense}{\ensuremath{\Sigma\tsb{dense}}\xspace}
\newcommand*{\Sigdiffuse}{\ensuremath{\Sigma\tsb{diffuse}}\xspace}
\newcommand*{\SigHi}{\ensuremath{\Sigma\sbHi}\xspace}
\newcommand*{\NHi}[1][]{\ensuremath{N\sbHi[#1]}\xspace}
\newcommand*{\nHi}[1][]{\ensuremath{n\sbHi[#1]}\xspace}
\newcommand*{\rhoHi}[1][]{\ensuremath{\rho\sbHi[#1]}\xspace}

\newcommand*{\hevapc}{\ensuremath{h\tsb{evap,c}}\xspace}
\newcommand*{\hpic}{\ensuremath{h\tsb{pi,c}}\xspace}
\newcommand*{\sigv}{\ensuremath{\sigma_v}\xspace}

\newcommand*{\sigturb}{\ensuremath{\sigma\tsb{turb}}\xspace}
\newcommand*{\pcgm}[1][]{\ensuremath{p\tsb{CGM\if\relax\detokenize{#1}\relax\else,#1\fi}}\xspace}
\newcommand*{\rhocgm}{\ensuremath{\rho\tsb{CGM}}\xspace}
\newcommand*{\Tcgm}{\ensuremath{T\tsb{CGM}}\xspace}
\newcommand*{\ncgm}[1][]{\ensuremath{n\tsb{%
      \if\relax\detokenize{#1}\relax%
      \else#1,\fi%
      CGM}}\xspace}
\newcommand*{\nHcgm}{\ncgm[H]\xspace}
\newcommand*{\mucgm}{\ensuremath{\mu\tsb{CGM}}\xspace}
\newcommand*{\cscgm}{\ensuremath{c\tsb{s,CGM}}\xspace}
\newcommand*{\hturb}{\ensuremath{h\tsb{turb}}\xspace}
\newcommand*{\hmix}{\ensuremath{h\tsb{mix}}\xspace}
\newcommand*{\hHi}{\ensuremath{h\sbHi}\xspace}
\newcommand*{\Tmix}{\ensuremath{T\tsb{mix}}\xspace}
\newcommand*{\sigmix}{\ensuremath{\sigma\tsb{mix}}\xspace}
\newcommand*{\mcloud}{\ensuremath{m\tsb{cloud}}\xspace}
\newcommand*{\ncloud}{\ensuremath{n\tsb{cloud}}\xspace}
\newcommand*{\sigcloud}{\ensuremath{\sigma\tsb{cloud}}\xspace}
\newcommand*{\rhocloud}{\ensuremath{\bar{\rho}\tsb{cloud}}\xspace}
\newcommand*{\tcool}{\ensuremath{t\tsb{cool}}\xspace}
\newcommand*{\tturb}{\ensuremath{t\tsb{turb}}\xspace}
\newcommand*{\CHi}{\ensuremath{C\sbHi}\xspace}
\newcommand*{\Da}{\ensuremath{\text{Da}}\xspace}
\newcommand*{\Leddy}{\ensuremath{L\tsb{eddy}}\xspace}
\newcommand*{\tilb}[1][]{\ensuremath{\tilde{b}%
    \if\relax\detokenize{#1}\relax%
    \else\tsb{#1}\fi}\xspace}
\newcommand*{\tilsig}[1][]{\ensuremath{\tilde{\sigma}%
    \if\relax\detokenize{#1}\relax%
    \else\tsb{#1}\fi}\xspace}

\newcommand*{\dd}{\mathrm{d}}
\newcommand*{\ee}{\mathrm{e}}
\makeatletter
\@ifpackageloaded{unicode-math}{%
  
}{%
  \iflatexml
    
  \else
    \IfPackageAtLeastTF{newtx}{2023/11/14}{%
      
    }{%
      
    }
  \fi
}\makeatother

\newlength{\myhalfimgsize}
\newcommand*\myplotone[1]{%
  \centering
  \leavevmode
  \includegraphics[width={%
        \ifdim\textwidth=\linewidth%
          2\myhalfimgsize%
        \else\ifdim\linewidth>\myhalfimgsize%
            \myhalfimgsize%
          \else%
            \linewidth
          \fi\fi}]{#1}%
}%

\newcommand*{\task}[1]{\textsc{#1}}
\newcommand*{\para}[1]{\texttt{#1}}

\DeclareSIUnit\yr{yr}
\DeclareSIUnit\Gyr{\giga\yr}
\DeclareSIUnit\Myr{\mega\yr}
\DeclareSIUnit\G{G}
\DeclareSIUnit\uG{\micro\G}
\DeclareSIUnit\pc{pc}
\DeclareSIUnit\Msun{M\ensuremath{_\odot}}
\DeclareSIUnit\kpc{\kilo\pc}
\DeclareSIUnit\Mpc{\mega\pc}
\DeclareSIUnit\dex{dex}
\DeclareSIUnit\deg{deg}
\DeclareSIUnit\erg{erg}
\DeclareSIUnit\mag{mag}
\DeclareSIUnit\Jy{Jy}
\DeclareSIUnit\mJy{\milli\Jy}
\DeclareSIUnit\beam{beam}
\DeclareSIUnit\bvla{beam_{VLA}}
\DeclareSIUnit\bfast{beam_{FAST}}
\DeclareSIUnit\pixel{pixel}

\newcommand*{\autoeqref}[1]{\hyperref[#1]{Equation~(\ref*{#1})}}

\iflatexml
  \providecommand{\textalpha}{\ensuremath{\alpha}}
\else
  \usepackage{textalpha}
\fi

\ProvideDocumentCommand\unit{om}{\si[#1]{#2}}
\ProvideDocumentCommand\qty{omm}{\SI[#1]{#2}{#3}}
\ProvideDocumentCommand\qtylist{omm}{\SIlist[#1]{#2}{#3}}
\ProvideDocumentCommand\qtyrange{ommm}{\SIrange[#1]{#2}{#3}{#4}}
\iflatexml\else
\fi

\graphicspath{{fig/}}

\defcitealias{2023ApJ...944..102W}{W23}
\defcitealias{2024ApJ...968...48W}{Paper~I}
\defcitealias{2024ApJ...973...15W}{Paper~II}
\defcitealias{2025ApJ...984...15Y}{Paper~IV}
\defcitealias{2019MNRAS.487..737J}{J19}
\defcitealias{2021MNRAS.502.3179T}{T21}

\shorttitle{Gas Accretion Implied by Diffuse \Hi Around M51}
\shortauthors{Lin et al.}

\received{2024 November 18}
\revised{2025 February 13}
\accepted{2025 February 14}
\published{\today}

\submitjournal{The Astrophysical Journal}

\begin{document}

\title{%
  FEASTS Combined with Interferometry.~III.
  The Low Column Density \Hi Around M51 and Possibility of Turbulent-mixing Gas Accretion
}

\correspondingauthor{\cjkname{Jing Wang}{王菁}}
\email{jwang\_astro@pku.edu.cn}


\author[0000-0002-4250-2709]{\cjkname{Xuchen Lin}{林旭辰}}
\affiliation{Department of Astronomy, School of Physics, Peking University, Beijing 100871, People's Republic of China}

\author[0000-0002-6593-8820]{\cjkname{Jing Wang}{王菁}}
\affiliation{Kavli Institute for Astronomy and Astrophysics, Peking University, Beijing 100871, People's Republic of China}

\author[0000-0002-8057-0294]{Lister Staveley-Smith}
\affiliation{International Centre for Radio Astronomy Research, University of Western Australia, 35 Stirling Highway, Crawley, WA 6009, Australia}
\affiliation{ARC Centre of Excellence for All-Sky Astrophysics in 3 Dimensions (ASTRO 3D), Australia}

\author[0000-0001-9658-0588]{\cjkname{Suoqing Ji}{季索清}}
\affiliation{Center for Astronomy and Astrophysics, Department of Physics, Fudan University, Shanghai 200438, People's Republic of China}
\affiliation{Key Laboratory of Nuclear Physics and Ion-beam Application (MOE), Fudan University, Shanghai 200433, People's Republic of China}

\author[0000-0002-5679-3447]{\cjkname{Dong Yang}{杨冬}}
\affiliation{Department of Astronomy, School of Physics, Peking University, Beijing 100871, People's Republic of China}

\author[0000-0002-5016-6901]{\cjkname{Xinkai Chen}{陈新凯}}
\affiliation{Kavli Institute for Astronomy and Astrophysics, Peking University, Beijing 100871, People's Republic of China}

\author[0000-0003-4793-7880]{Fabian Walter}
\affiliation{Max-Planck-Institut f\"ur Astronomie, K\"onigstuhl 17, D-69117, Heidelberg, Germany}

\author[0000-0001-8813-4182]{Hsiao-Wen Chen}
\affiliation{Department of Astronomy and Astrophysics, The University of Chicago, 5640 S. Ellis Avenue, Chicago, IL 60637, USA}

\author[0000-0001-6947-5846]{Luis C. Ho}
\affiliation{Kavli Institute for Astronomy and Astrophysics, Peking University, Beijing 100871, People's Republic of China}

\author[0009-0003-3527-8520]{\cjkname{Peng Jiang}{姜鹏}}
\affiliation{National Astronomical Observatories, Chinese Academy of Sciences, 20A Datun Road, Chaoyang District, Beijing 100101, People's Republic of China}

\author[0000-0001-8057-5880]{Nir Mandelker}
\affiliation{Centre for Astrophysics and Planetary Science, Racah Institute of Physics, The Hebrew University, Jerusalem 91904, Israel}

\author[0000-0002-8379-0604]{Se-Heon Oh}
\affiliation{Department of Physics and Astronomy, Sejong University, 209 Neungdong-ro, Gwangjin-gu, Seoul, Republic of Korea}

\author[0000-0002-1605-0032]{Bo Peng}
\affiliation{Max-Planck Institut f{\"u}r Astrophysik, Karl-Schwarzschild-Stra\ss{}e 1, D-85741 Garching, Germany}

\author[0000-0002-4288-599X]{C\'eline P\'eroux}
\affiliation{European Southern Observatory, Karl-Schwarzschild-Stra\ss{}e 2, D-85748 Garching, Germany}
\affiliation{Aix Marseille Universit\'e, CNRS, LAM (Laboratoire d'Astrophysique de Marseille) UMR 7326, F-13388 Marseille, France}

\author[0000-0002-2941-646X]{\cjkname{Zhijie Qu}{屈稚杰}}
\affiliation{Department of Astronomy and Astrophysics, The University of Chicago, 5640 S. Ellis Avenue, Chicago, IL 60637, USA}
\affiliation{Department of Astronomy, Tsinghua University, Beijing 100084, People's Republic of China}

\author[0000-0002-9279-4041]{Q. Daniel Wang}
\affiliation{Department of Astronomy, University of Massachusetts, Amherst, MA 01003, USA}

\begin{abstract}
  With a new joint-deconvolution pipeline, we combine the single-dish and interferometric atomic hydrogen (\Hi) data of M51 observed by the Five-hundred-meter Aperture Spherical radio Telescope (FAST) (FEASTS program) and the Very Large Array (VLA) (THINGS)\@.
  The product data cube has a typical line width of \qty{13}{\km\per\s} and a $2\sigma$ line-of-sight (\los) sensitivity of \Hi column density $\NHi\sim\qtyNHilim[]$ at a spatial resolution of \qtyVLAfwhm (\qtyphysVLAfwhm).
  Among the \Hi-detected \los{}s extending to \qty{\sim50}{\kpc}, \qty{\sim89}{\percent} consist of diffuse \Hi only, which is missed by previous VLA observations.
  The distribution of dense \Hi is reproduced by previous hydrodynamical simulations of this system, but the diffuse component is not, likely due to unresolved physics related to the interaction between the circumgalactic and interstellar media.
  With simple models, we find that these low-\NHi structures could survive the background ultraviolet photoionization, but are susceptible to the thermal evaporation.
  We find a positive correlation between \los velocity dispersion (\sigv) and \NHi with a logarithmic index of \num{\sim0.5}.
  Based on existing turbulent mixing layer (TML) theories and simulations, we propose a scenario of hot gas cooling and accreting onto the disk through a TML, which could reproduce the observed power index of \num{\sim0.5}.
  We estimate the related cooling and accretion rates to be roughly one-third to two-thirds of the star-formation rate.
  A typical column density of diffuse \Hi (\qty{\sim1e19}{\per\cm\squared}) can be accreted within \qty{300}{\Myr}, the interaction timescale previously estimated for the system.
  Such a gas accretion channel has been overlooked before, and may be important for gas-rich interacting systems and for high-redshift galaxy evolution.
\end{abstract}

\section{Introduction}
\label{sec:intro}
The evolution of galaxies is largely determined by the cosmic baryon cycle \citep{2020ARA&A..58..363P}, which involves complex hydrodynamic processes \citep{2023ARA&A..61..473C}.
In particular, the multiphase nature of gas poses challenges to both theoretical and observational understanding of galaxy evolution \citep{2017ARA&A..55...59N}.
The accretion and depletion of neutral gas regulate the supply of star-forming material, thereby influencing the growth of stellar mass and the formation of disk structures on scales of \qtyrange{1}{10}{\kpc} \citep{1998MNRAS.295..319M}.
Continuous gas accretion into the interstellar medium (ISM) (the gas inside galaxy disks) is necessary because the gas within star-forming regions would otherwise be exhausted within a few gigayears \citep{2017ApJS..233...22S,2020ARA&A..58..363P,2020ARA&A..58..157T}.

In theory, gas accretion can occur through various modes, including the cosmological cold- and hot-mode accretion, the redistribution of satellite gas, and the galactic fountain \citep{2017ASSL..430....1P,2023ARA&A..61..131F}.
Recent multiwavelength observations have revealed that gas is accreted primarily from the intergalactic medium (IGM) (the gas outside dark matter halos) in an ionized phase, while the cooling and condensation into a neutral phase happen later within the circumgalactic medium (CGM) (the gas between IGM and ISM) \citep{2017ASSL..430....1P}.
Possible mechanisms happening during gas accretion and cooling include the precipitation from thermal instability \citep[e.g.,][]{2017ApJ...845...80V,2018ApJ...854..167G}, the condensation by mixing \citep{2020MNRAS.492.1970G}, and the rotating cooling flow \citep{2022MNRAS.514.5056H,2024MNRAS.530.1711S}.
However, the details and relative importance of these physical processes remain uncertain \citep{2023ARA&A..61..131F}.
Observationally, though not directly tracing the material accreted from the IGM, the neutral gas in CGM serves as an indicator of high-density, low-temperature regions, providing insights into the overall efficiency of multiphase gas accretion \citep{2017ASSL..430...15R,2020MNRAS.498.2391N}.

Due to their low column density, cold (\qty{<1e4}{\K}) gas components within the CGM have been primarily studied using absorption lines.
Statistically, these absorption-line detections sample a large area around the associated galaxies, even extending beyond the virial radius \citep[e.g.,][]{2020ApJ...900....9L,2021ApJ...912....9W,2023ApJ...944..101B}, in line with the large-scale structures traced by neutral gas in simulations \citep[e.g.,][]{2019MNRAS.482L..85V,2023MNRAS.518.5754R}.
However, the current sample size is not large enough to fully explore the dependence of absorption-line detections on diverse galaxy properties and other factors, such as the group environment and tidal interactions \citep{2020MNRAS.497..498C,2022MNRAS.516.5618P,2023ApJ...944..101B}.
One example is the dependence on the azimuthal angle of absorbers around galaxies, which is predicted by theories \citep{2020MNRAS.499.2462P} but still inconclusive in observations \citep{2011ApJ...743...10B,2015ApJ...815...22K,2023MNRAS.523..676W}.
The geometric degeneracies and the incomplete galaxy optical information complicate the association of absorption systems with galaxies, and further make it difficult to interpret their distances and relative velocities \citep{2019ApJ...878...84M}.
Near the disk--CGM interface, the saturation of Lyman~\textalpha\ absorption lines at intermediate neutral atomic hydrogen (\Hi) column densities ($\NHi\sim\qtyrange{1e17}{1e19}{\per\cm\squared}$) and the limited availability of high-quality Lyman-limit system observations pose challenges to understanding the ISM--CGM connection \citep{2023ApJ...944..101B,2023ApJ...954..115K,2023MNRAS.523..676W}.

Directly mapping \Hi around galaxies across a wide range of column densities is essential to solving many of the aforementioned challenges and uncertainties.
The \qty{21}{\cm}-emission \Hi mapping offers a wide field of view around galaxies, covering the various conditions and environments where multiphase gas resides.
If the sensitivity is sufficient, it can directly trace the final stages of transition into \Hi and accretion from the CGM onto the ISM disk.
The CGM--ISM interface is where matter, energy, and momentum are actively exchanged and recycled, driven by processes like thermal conduction, radiative cooling, and turbulent mixing \citep{2023ARA&A..61..131F}, which could be powerfully constrained with the direct mapping of \Hi column densities and velocities.
In addition, direct mapping is not limited by the incidence of quasars, making the process of sample selection much simpler than absorption studies.
The direct mapping would also help to refine simulations that have difficulties in linking the cosmological accretion history with small-scale gas physics \citep{2020NatRP...2...42V,2023ARA&A..61..131F}.

With the high sensitivity of the Five-hundred-meter Aperture Spherical radio Telescope (FAST), the FAST Extended Atlas of Selected Targets Survey \citep[FEASTS;][hereafter \citetalias{2023ApJ...944..102W}]{2023ApJ...944..102W}%
\footnote{
  Data are publicly available at \url{https://github.com/FEASTS/LVgal/wiki}.
}
is able to map the \Hi in the local Universe down to a $3\sigma$ column density limit of $\NHi\sim\qty{5e17}{\per\cm\squared}$, nearly reaching the \NHi range of Lyman-limit systems.
\citetalias{2023ApJ...944..102W} detected large-scale low-\NHi gas associated with the NGC~4631 group, showing the potential of FAST to bridge the gap between ISM and CGM studies.
With a larger sample of 11 systems, a follow-up study by \citet[hereafter \citetalias{2024ApJ...968...48W}]{2024ApJ...968...48W} estimated the \Hi number density based on the assumption of pressure balance with the CGM\@.
They found that the low-density, diffuse component of \Hi missed by interferometers is ubiquitous, and that the diffuse fraction increases with the strength of tidal interactions.
This low-density faint \Hi is likely a direct result of CGM--ISM material exchange, potentially enhanced by tidal interactions.

Tidally interacting galaxies offer unique laboratories for studying the physical processes connecting the CGM and ISM in the context of both galaxy evolution and multiphase gas dynamics.
Observational and theoretical studies have found that intense tidal interactions during mergers could enhance star formation \citep{2019ApJ...881..119P,2022MNRAS.517L..92E}, and at the same time, the gas mass remains constant \citep{2021MNRAS.504.1888Q,2022ApJ...934..114Y,2023MNRAS.523..720L} or even increases \citep{2015MNRAS.448..221E,2018MNRAS.478.3447E,2019MNRAS.485.1320M,2019ApJ...870..104S} with possibly diluted metallicity \citep{2019MNRAS.482L..55T};
simulated and observed major-merger remnants also exhibit rejuvenation \citep{2017MNRAS.470.3946S}.
These findings suggest rapid evolution in galaxy mass assembly, driven by active gas cooling, inflow, and consumption at various merger stages.
Mergers are known to drag large and long-lived tidal tail structures \citep[e.g.,][]{2007ApJ...659L.115C,2021ApJ...922L..21Z}, enlarging the interface between cold \Hi structures and the hot CGM and possibly catalyzing gas exchange between them.
The widespread morphology, perturbed kinematics, and rapid evolution of merging systems offer a wide range of gas conditions, making them ideal for testing gas flow-related physical scenarios.
These systems may also serve as analogs for galaxies at higher redshifts, where gas-rich mergers are frequent on average \citep{2014ARA&A..52..291C,2020ARA&A..58..363P,2024arXiv240709472D}, and are likely a significant channel of gas accretion.

\object{M51a} (NGC~5194, aka the Whirlpool Galaxy) is in the late stage of merging with \object{M51b} (NGC~5195), as supported by simulations reproducing the interferometrically observed \Hi distribution and kinematics \citep{1990AJ.....99.1798H,2003Ap&SS.284..495T,2010MNRAS.403..625D}.
Strong tidal effects induced the formation of evident spiral arms and an extended tail, observed in both stellar light \citep{1850RSPT..140..499R} and gas emission \citep{1990AJ....100..387R,2013ApJ...779...43P}.
Previous studies have mostly focused on the inner disk of M51a \citep[e.g.,][]{2014ApJ...784....4C,2014PASJ...66...77O,2023ApJ...944...86M}, where the galaxy's spiral structure is rather regular and of grand design, and multiwavelength data are abundant.
Recent FAST observations have revealed extended low-\NHi gas surrounding the pair \citetext{\citetalias{2024ApJ...968...48W}; \citealp[see also][]{2023MNRAS.521.2719Y}}, especially at the unexpected west and south sides.
Compared to other galaxies that FEASTS has already observed \citepalias{2024ApJ...968...48W}, the faint \Hi in M51 is not closely attached to the bright and dense \Hi detected in interferometric observations \citep{1990AJ....100..387R,2008AJ....136.2563W}, but extends much further (see \autoref{fig:SigHI_prof} in \autoref{app:sec:FEASTS_NHi}).
Without nearby satellites, this large extension of relatively diffuse \Hi gas likely originates from the strong tidal interaction.
The diffuse \Hi may represent an intermediate phase between the colder gas and hot CGM, and in this study, we explore its role in the evolution of this system.

Despite the high sensitivity, FAST is limited by its \ang{;\sim3;} spatial resolution due to its single-dish nature.
One long-standing goal in radio astronomy is to combine low-resolution single-dish \Hi data with high-resolution interferometric data, which usually miss total flux because of the short-spacing problem \citep{2002ASPC..278..375S,2023PASP..135c4501P}.
Among the many possible methods, jointly deconvolving the interferometric dirty cube and single-dish \Hi data is the one most likely to coherently recover both the spatial and kinematic information, but has rarely been applied beyond the Local Group \citep[e.g.,][]{1988A&A...202..316C} due to many difficulties.
Especially, the apparent size of \Hi structures decreases significantly at large distances, making them difficult to resolve with single-dish telescopes that have either large beams (e.g., the Green Bank Telescope) or significant side lobes \citep[e.g., Arecibo,][]{2005AJ....130.2598G}.

This series of studies focuses on combining FEASTS with interferometric data, leveraging FAST's relatively small beam and low side-lobe level.
In the first two papers \citetext{\citetalias{2024ApJ...968...48W}; \citealp[hereafter \citetalias{2024ApJ...973...15W}]{2024ApJ...973...15W}}, the cleaned interferometric models of 11 galaxies are compared or linearly combined with their FAST cubes, which contains the diffuse \Hi missed by interferometers.
The diffuse \Hi was found to rotate more slowly and have a higher velocity dispersion than the dense one, possibly forming a thick layer extending vertically into the CGM\@.
In this paper (Paper~III), we conduct a detailed study of M51, tackling its unexpectedly extended \Hi structure.
We develop a new pipeline using the maximum entropy method \citep[MEM, e.g.,][]{1978Natur.272..686G} to jointly deconvolve the FAST cube and the Very Large Array (VLA) data obtained by The \Hi Nearby Galaxy Survey \citep[THINGS,][]{2008AJ....136.2563W}, simultaneously achieving the deep sensitivity and high resolution of the two data sets.
\citetalias{2025ApJ...984...15Y} \citep{2025ApJ...984...15Y} presents a statistical study of the extraplanar \Hi around edge-on galaxies.

This paper is organized as follows.
The data sets and analysis methods are introduced in Sections \ref{sec:sample} and~\ref{sec:analyses}, including the deconvolution pipeline and decomposition of diffuse gas components.
We present the results in \autoref{sec:results}.
The velocity dispersion is compared pixelwise with column density, and the \Hi number density and thickness are calculated assuming pressure balance with the CGM\@.
In \autoref{sec:tml_cool}, we discuss the possible mechanisms of the \Hi cooling and accretion driven by turbulent mixing, and summarize in \autoref{sec:summary}.
Throughout this paper, the variables with a subscript of ``\Hi'' include only the contribution from hydrogen, and a factor of \num{1.4} is applied when helium needs be considered.
Meanwhile in \autoref{sec:tml_cool}, the density of CGM, and thus the cooling rate, takes the helium into account so that we can compare them with the star-formation rate (SFR)\@.

\section{Data}
\label{sec:sample}
We take the coordinates and distance of M51a as those adopted by THINGS, $\alpha_{2000}=\ang{202.4696;;}$, $\delta_{2000}=\ang{47.1952;;}$ \citep{1994ApJ...421..122T} and $D=\qty{8.0}{\Mpc}$ \citep{2004AJ....127.2031K}.
Its stellar mass (\qty[parse-numbers=false]{10^{10.73}}{\Msun}) and SFR (\qtysfr) are taken from the $z=0$ Multiwavelength Galaxy Synthesis \citep[z0MGS;][]{2019ApJS..244...24L} as in \citetalias{2024ApJ...968...48W}, and the $B$-band optical size $\Rzs=\qtyRzs$ (\ang{;5.0;}) \citep{2003A&A...412...45P}.
We calculate its virial mass \Mvir from the stellar mass, using the relation of \citet{2010ApJ...717..379B}.
Here, we do not use the dynamical mass (\qty[parse-numbers=false]{10^{12.190}}{\Msun}) from \citet{2017ApJ...843...16K} as in \citetalias{2023ApJ...944..102W}, because for M51, they detected a small number of group members (only six).
They estimated a mass of \qty[parse-numbers=false]{10^{12.685}}{\Msun} from the $K$-band luminosity, and the value we use serves as a middle value.
The coordinates of M51b are taken from the NASA/IPAC Extragalactic Database (NED), $\alpha_{2000}=\ang{202.4983;;}$ and $\delta_{2000}=\ang{47.2661;;}$ \citep{2006AJ....131.1163S}.

\subsection{FAST Data}
The FAST observations of the FEASTS sample were carried out using basket-weaving on-the-fly scans.
The data reduction pipeline implements the standard procedures of single-dish \Hi data processing, and is optimized for the FEASTS data \citepalias{2023ApJ...944..102W}.
The \Hi data cubes have a beam of \ang{;3.24;} FWHM, and a channel width of \qty{1.610}{\km\per\s}.
The FEASTS observations of M51 reach an rms level of \qty{0.80}{\mJy\per\bfast}, corresponding to a $3\sigma$ column density detection limit of \qty{4.0e17}{\per\cm\squared}, assuming a line width of \qty{20}{\km\per\s}.
More details of the observations of M51 can be found in \citetalias{2024ApJ...968...48W}.

\subsection{VLA Data}
\label{ssec:vla_data}
The main interferometric data we use are the calibrated and continuum-removed visibility obtained by the THINGS \citep{2008AJ....136.2563W}, which covers a heliocentric radial velocity range between \qtylist{214.9;704.5}{\km\per\s} with a channel width of \qtychanwidth (96 channels).
The continuum subtraction of the THINGS visibility is improved as described in \autoref{app:sec:contsub}.
We verify that the contribution of the $w$-components \citep{2017isra.book.....T} is insignificant in this visibility data set.

The naturally weighted dirty cube is generated using MIRIAD task \task{invert}, during which we apply an $\text{FWHM}=\ang{;;3}$ Gaussian taper to increase the signal-to-noise ratio (S/N)\@.
The pixel size is the same as that of THINGS products, \ang{;;1.5}, and the image size is $\num{1800}\times\num{1800}$ pixels ($\ang{;45;}\times\ang{;45;}$).
The dirty cube has a noise level of $\sigint=\qty{0.379}{\mJy\per\bvla}$, corresponding to a $3\sigma$ column density limit of \qty{4.6e19}{\per\cm\squared}, assuming a line width of \qty{20}{\km\per\s}.
The corresponding clean beam has a shape of \VLAcleanbeam with position angle of $\PA=\ang{67.52;;}$, or a physical resolution of \qtyphysVLAfwhm.

\begin{figure*}
  \includegraphics{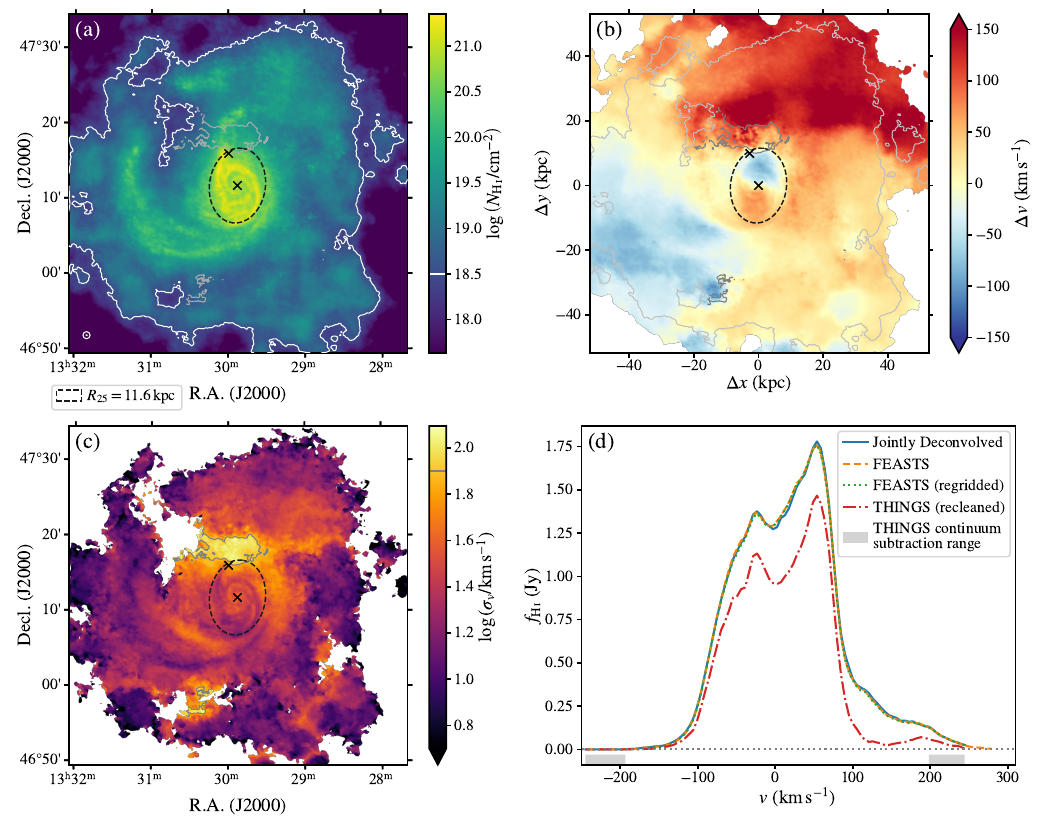}
  \caption{
    The moment maps (a--c) and the spectrum (d) of the jointly deconvolved data.
    The centers of M51a and M51b are marked as black crosses in panels (a)--(c).
    The stellar disk size ($\Rzs=\qtyRzs$) is indicated by the black ellipse, with $\PA=\ang{172;;}$ and $i=\ang{42;;}$ \citep[as adopted by THINGS]{2008AJ....136.2872T}.
    The gray contour encloses the region with a moment-2 value higher than \rnghighlydisperse, and the white contour indicates the $2\sigma$-confidence \los column density limit, \qtyNHilim[\sim].
    In the lower left corner of panel~(a), the open white circle displays the diameter of the instrumental resolution (\qtyFWHMavg) at $\NHi\sim\qtyNHilim[]$, and the white ellipse within is the restoring beam (\VLAcleanbeam).
    The actual resolution transits between these two limits, reaching the best one when $\NHi>\qty{1.6e19}{\per\cm\squared}$.
    In panel~(d), the spectrum of the jointly deconvolved cube (blue solid) is compared with those of the single-dish FEASTS (orange dashed) and VLA THINGS (red dashed--dotted; recleaned).
    As reference, the spectrum of the FEASTS cube after regridded into the interferometric coordinates is shown as the green dotted line.
    The gray shades indicate the channel range of continuum subtraction for THINGS data.
    The central velocity of $\vsys=\qtyvcent$ is taken from the velocity modeling (see \autoref{ssec:vfieldmod}).
  }
  \label{fig:deconv_result}
\end{figure*}

In \autoref{fig:deconv_result}(d), we plot the cleaned spectrum of this corrected THINGS visibility as red dashed--dotted line.
Using MIRIAD \task{clean}, we largely follow the procedures of \citet{2008AJ....136.2563W}, including the rescaling of residual, with a cleaning threshold of $1.5\sigma$.

\section{Analyses}
\label{sec:analyses}
\subsection{Jointly Deconvolving the FAST and VLA Cubes}
\label{ssec:pipeline}
We developed a pipeline to jointly deconvolve the VLA dirty cube and FAST cube per channel using MEM\@.
MEM seeks the positive solution model image \modim ($j$ denotes the serial number of pixels) that simultaneously
\begin{enumerate}
  \item \label{mosmem:idea:chi2}
        fits the input single-dish or dirty image, i.e., the rms value of residual map $r_j$ is smaller than a threshold \thres{}, and
  \item \label{mosmem:idea:default}
        has the maximum information entropy
        \begin{equation}
          H = -\sum_j \modim \ln\frac{\modim}{\ee\defim}
          = -\sum_j \modim \left(\ln\frac{\modim}{\defim} - 1\right)
        \end{equation}
        relative to an input reference image \defim, which is referred to as ``a~priori image'' by \citet{1985A&A...143...77C} and as ``default image'' in the MIRIAD task \task{mosmem} \citep{1990ApJ...354L..61S}.
        Traditionally, this reference image is instead set to be a flat image with the total flux of the single-dish image.
\end{enumerate}
Condition~\ref{mosmem:idea:chi2} is the basic chi-square requirement for a possible solution, and condition~\ref{mosmem:idea:default} forces the model to trace the reference image \defim when the constraint from condition~\ref{mosmem:idea:chi2} is weak.
In this pipeline, we use the MIRIAD task \task{mosmem} \citep{1985A&A...143...77C,1988A&A...202..316C,1990ApJ...354L..61S,1996A&AS..120..375S}, a common implementation of MEM deconvolution for radio astronomy.

Unlike traditional MEM applications \citep[see the summary by][]{2002ASPC..278..375S}, this new pipeline has these two characteristics:
\begin{enumerate}
  \item A non-flat reference cube \defim is passed to the MEM deconvolution, along with the interferometric dirty cube \intim and the single-dish cube \sdim.
  \item We \emph{iteratively} invoke the MEM deconvolution.
        The reference \defim is initially set to be \sdim, and is adaptively adjusted between each iteration according to the output model \modim and residuals.
\end{enumerate}
By this pipeline, we more exhaustively exploit the available information in the data, and reduce and flatten the residuals more thoroughly than before.
The noise level of the jointly deconvolved result is expected to be between the interferometric (\sigint) and single-dish (\sigsd) noise levels, and we estimate it as $\sqrt{\sigint\sigsd}$.
The step-by-step technical details of the pipeline are given in \autoref{app:sec:pipeline} for readers who are interested.

The output model \modim is convolved with a restoring beam chosen to be the VLA clean beam (\VLAcleanbeam), without adding back either of the two residuals.
\autoref{fig:deconv_result} gives the moment maps and spectrum of the jointly deconvolved M51 \Hi data.
No cube mask is applied except when we calculate the moment-2 map, which therefore appears smaller.
The reason is that the line-of-sight (\los) velocity dispersion measurement is more sensitive to noise than the column density is.
The cube mask is generated with SoFiA~2 \citep{2015MNRAS.448.1922S,2021MNRAS.506.3962W} from the model cube plus a Gaussian noise of $\sqrt{\sigint\sigsd}$ at the resolution of restoring beam.

Our pipeline is justified by the mock tests in \autoref{app:ssec:recovery}, which show that the accuracy of moment-2 map increases significantly with a SoFiA mask, and that at the resolution of restoring beam (\qtyVLAfwhm, or \qtyphysVLAfwhm), the $2\sigma$ column density limit is \qtyNHilim for a single \los (white contour in \autoref{fig:deconv_result}a).
This \NHi limit is roughly the geometric mean of the sensitivities of THINGS and FEASTS (\autoref{sec:sample}), justifying our choice of $\sqrt{\sigint\sigsd}$ as the estimated noise level.
In \autoref{app:ssec:noise_spec}, we also model the instrumental resolution of the jointly deconvolved cube at a given \NHi level using the noise Fourier spectra of THINGS and FEASTS data.
We find that the instrumental resolution at the above \NHi limit (\qtyNHilim) is \qtyFWHMavg (\qtyphysFWHMavg[]), and that a resolution of restoring beam (\qtyVLAfwhm) can be reached at \qty[input-comparators=\gtrsim]{\gtrsim1.6e19}{\per\cm\squared}.
The mock-verified effective resolution (\qtyVLAfwhm) is much better than the instrumental one (\qtyFWHMavg), likely because the intrinsic spatial scale for the \Hi gas at \qtyNHilim is much larger than the instrumental resolution.
In \autoref{fig:deconv_result}(a), we plot at the lower left corner both the restoring beam and a circular \qtyFWHMavg[] beam as the resolution reference.

We define our \emph{\los sample} as the \los{}s with a finite moment-2 measurement (\sigv, the \los velocity dispersion) smaller than \rnghighlydisperse, i.e., the colored regions out of the gray contour in \autoref{fig:deconv_result}(c).
We set the upper limit as \qty{80}{\km\per\s} because these high-\sigv regions may be related to shocks or the projection of multiple components in the northern region, where M51 encounters M51b.

The total \Hi mass we jointly deconvolved is \qty[parse-numbers=false]{10^{9.59}}{\Msun}, the same as the FEASTS value \citepalias{2024ApJ...968...48W} and \num{1.55} times the THINGS-only value.
We largely recovered the extra diffuse \Hi flux detected by FAST relative to the THINGS data (\autoref{fig:deconv_result}d).
In \autoref{app:ssec:mom0_comp}, we compare the moment-0 map from the joint deconvolution with those of FEASTS, recleaned THINGS cube, and the result of linear combination.
The channel maps of the model and residual cubes can be found in \autoref{app:sec:chan_map}.
They consistently show that the joint deconvolution pipeline produces a data cube with smooth transition from high- to low-density regions.

\subsection{Removing Projection Effects from the Velocity Field}
\label{ssec:vfieldmod}
In later analyses of gas entrainment, one key parameter to be used is the velocity of \Hi moving in the CGM, which is reflected in the moment-1 map (\autoref{fig:deconv_result}b) but suffers from projection effects.
The velocity fields both within and beyond \Rzs show dipolar and spider-like morphology, indicating the possibility of partially removing the projection effects by modeling the velocity field with tilted rings, as done previously in other large-scale interacting systems \citep[e.g.,][]{2019MNRAS.486..504S}.
The kinematics of a tidally perturbed system cannot be well described by rings, and the outer \Hi structure often does not consist of full rings (see \autoref{ssec:tail}), but to the first order, the modeling provides a best-effort correction for projection.

With \emph{BBarolo} (or \textsuperscript{3D}Barolo) \citep{2015MNRAS.451.3021D} 3DFIT task, we conduct the fitting in four rounds before obtaining the final results.
Each model consists of \num{130} rings with a radial increment of \ang{;;10}.
In the first round, these rings are divided into three groups with a radial range of \ang{;;<300}, \ang{;;300}--\ang{;;570}, and \ang{;;>570}, ensuring relatively constant \PA's within each group.
These three groups roughly correspond to three visually identified regions with distinct velocity patterns: the inner disk, the transitional region, and outer tidal structures.
We fit the rings in each group separately, and in later rounds, we cancel the grouping.
Results of each round are fixed or used as the initial guess for the next round.
In the first round, the initial values of central position, systemic velocity (\vsys), and inclination ($i$) take the kinematic value used by THINGS \citep{2008AJ....136.2872T}, while those of rotational velocity (\vrot) and \PA are visually guessed.
The radial velocity \vrad is set as zero initially.
The variation ranges of $i$ and \PA are \ang{30;;} and \ang{60;;} in the first two rounds, and decrease to \ang{5;;} and \ang{15;;} in the third round.
The ring centers and \vsys are fixed after the first round.
In the final round, \vrot and \vrad are fitted with all other parameters fixed.
By doing so, we gradually tighten the constraints on the model toward the final results.

\begin{figure}
  \centering
  \includegraphics{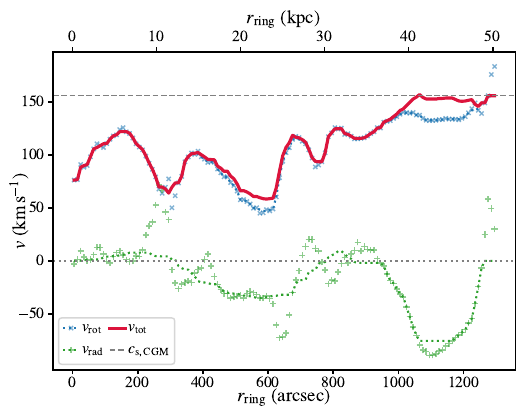}
  \caption{
    The profile of fitted rotational (\vrot), radial (\vrad), and total (\vtot) velocities from the tilted-ring modeling using \emph{BBarolo}.
    The value of \vtot (thick red solid line) is calculated as $\sqrt{\vrot^2+\vrad^2}$.
    The crosses are the fitted values of \vrot (blue) and \vrad (green), and the dotted lines are the median-filtered profiles;
    the filter sizes are \ang{;;50} and \ang{;;210} for \vrot and \vrad, respectively.
    By convention, a positive \vrad in \emph{BBarolo} indicates inflowing for a galaxy rotating counterclockwise as M51a.
    The value of the sound speed in CGM ($\cscgm=\qtycscgm$) is indicated by the horizontal gray dashed line.
  }
  \label{fig:ring_model}
\end{figure}

Two parameters crucial to our later analyses are the total velocity $\vtot=\sqrt{\vrot^2+\vrad^2}$ and the inclination $i$.
In \autoref{fig:ring_model} are the fitted velocity profiles, and the full fitting results are presented in \autoref{app:sec:vfieldmap}.
The fluctuation reflects the complexity of this system and the degeneracy between velocities and geometric parameters.
To suppress the fluctuation, we smooth the profiles of \vrot and \vrad with median filters of \ang{;;50} and \ang{;;210} wide, respectively.
The \vtot profile (red solid line in \autoref{fig:ring_model}) is calculated from the smoothed profiles of \vrot and \vrad (dotted lines).
We construct spatial maps of \vtot and $i$ based on the tilted-ring model for later pixelwise analyses.
The details of the map construction are given in \autoref{app:sec:vfieldmap}.

\subsection{Separating Diffuse and Dense \Hi}
\label{ssec:bigauss_decomp}
One advantage of the M51 system is that a large fraction of \Hi-detected \los{}s have a low \Hi column density.
These \los{}s are likely to be mainly composed of diffuse \Hi.
Physically, we expect the diffuse \Hi to have relatively low number densities, high velocity dispersions, high thermal temperatures, and large variance scales.
Its properties should lie between those of the CGM hot gas and the ``dense'' \Hi typically captured by interferometry \citepalias{2023ApJ...944..102W}.
Its temperature may be of \qty{\sim5e3}{\K}, the typical value of the warm neutral medium detected in the Milky Way \citep[MW;][]{2023ARA&A..61...19M}, so as to roughly remain neutral and thermally stable \citep{1969ApJ...155L.149F}.

Because the data have a limited spatial resolution, the observed diffuse \Hi possibly includes unresolved small-scale dense \Hi cloudlets \citep{2020MNRAS.494L..27G}.
The observation thus provides an ensemble measure of gas properties \citep{2018ApJ...854..167G}, meaning that instead of dealing with the behavior of individual cloudlets, we focus on the collective properties and behavior of the gas.
The unresolved dense \Hi cloudlets either are insignificant in mass compared to the diffuse \Hi, or else behave as an ensemble like the diffuse \Hi (like cloudlets in a fog/mist).
While the large-scale \Hi distribution and motion in the M51 system should be effectively shaped by the tidal interaction (but see \autoref{ssec:tail}), the localized survival of \Hi depends on the interaction with the immediate radiative and thermodynamic environment of the CGM (in the off-plane, less self-shielded direction).
This environment leaves signatures in the ensemble \Hi properties, and can be compared with the prediction of simple models in the assumption of quasi-equilibrium.

With these considerations, we develop a simple procedure to separate the \los \Hi spectrum into the broad and narrow components, which corresponds to the more diffuse (warmer) and denser (cooler) gas parts.
We select bright channels along each \los, and fit them with one or two Gaussian component(s) plus a flat positive baseline.
These Gaussian models are relatively arbitrarily required to have $\sigv\lesssim\qty{35}{\km\per\s}$ and peak intensity above $2\sigint$.
Two Gaussian components are used only when two distinct peaks above $1.5\sigint$ are identified;
otherwise, only one component is used.
The detailed procedure is given in \autoref{app:sec:bigauss_decomp}.

We separate the \Hi into the following three components, and refer to them as
\begin{enumerate}
  \item \emph{Diffuse-only \Hi.}
        No fitting is conducted in these \los{}s.
        The spectral peak is smaller than $2\sigint$, corresponding to a maximum \NHi of $\qty{\sim9.5e19}{\per\cm\squared}\times(\sigv/\qty{18}{\km\per\s})$.
  \item \emph{Dense \Hi} refers to the fitted Gaussian components.
        For the \los{} spectra with two dense components (only \qty{7}{\percent} of the dense-\Hi--bearing \los{}s), the one with a higher column density is denoted as the \emph{main} component, and the other is the \emph{secondary} one.
  \item \emph{Dense-associated diffuse \Hi} is the residual of the fitted dense \Hi component.
        This component and the diffuse-only \Hi are referred to as \emph{diffuse \Hi} together.
\end{enumerate}

Among the LOS sample, \qty{29}{\percent} of the mass is diffuse-only, \qty{20}{\percent} is dense-associated diffuse, and \qty{51}{\percent} is dense.
In the same sample, \qty{\sim89}{\percent} of the \los{}s are associated with diffuse-only \Hi.
Because this study focuses on \los properties, the spectral decomposition's goal is equivalently reduced to simply separating the diffuse-only \los from other \los{}s.
This simple goal mitigates the uncertainties coming from the  procedure.

In the following, the \los{}s of diffuse-only \Hi are used as the fiducial sample for discussion, while the dense-associated diffuse \Hi is only shortly discussed in Sections \ref{ssec:localized} and~\ref{ssec:radial_property} for the benefit of completeness and consistency check.
In the future, when extending this study to a sample of galaxies, we will conduct more-careful \los decomposition using tools like \emph{BAYGAUD} \citep{2022ApJ...928..177O}.

\section{Results}
\label{sec:results}
We present the results of this paper in this section.
We first present the global kinematics and morphologies of the \Hi components that we separated, especially the \Hi in the diffuse-only region.
We then focus on the statistical distributions of the \los parameters, particularly those related to the diffuse \Hi, including the relation between the velocity dispersion and \Hi column density, and the radial profiles of the number density and \Hi thickness.
These relations reflect the physical processes that regulate these parameters, such as the heating or cooling of diffuse \Hi.

\subsection{Inspecting the Moment Maps of \Hi in the Inner Region}
\label{ssec:decompose_result}
\begin{figure*}
  \centering
  \includegraphics{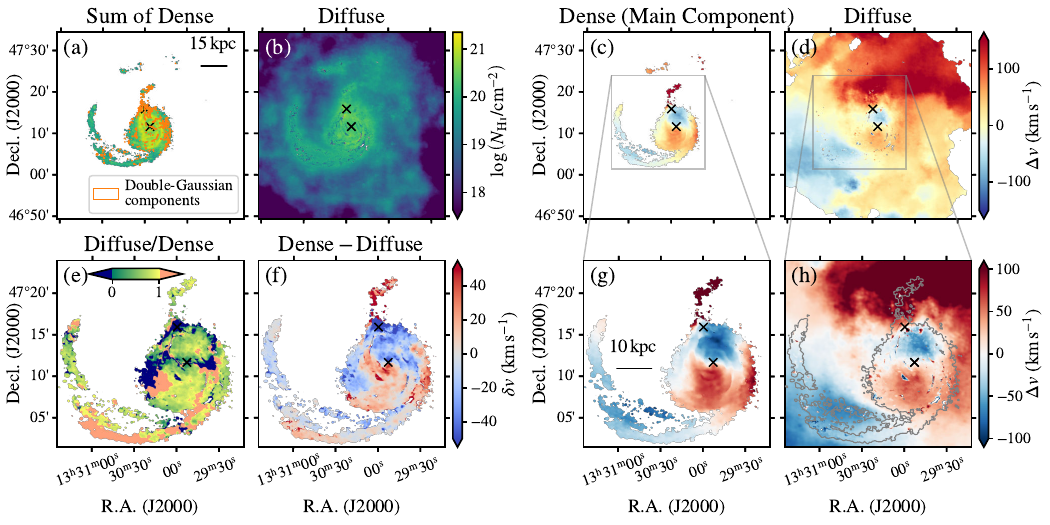}
  \caption{
    The moment-0 and -1 map properties of the dense and diffuse \Hi.
    The locations of M51a and~b are marked as crosses.
    (a)~The moment-0 map of total dense \Hi.
    The \los that consists of two dense components is enclosed by orange contours.
    (b)~The moment-0 map of diffuse \Hi.
    The negative values from noise are blanked.
    (c)~and (d)~The moment-1 maps of the main dense \Hi component and the diffuse \Hi.
    The zoomed-in version is in panels (g) and~(h).
    The dense-associated diffuse \Hi is enclosed by the gray contour in panel~(f).
    (e)~The ratio of \los velocity, relative to the central systemic velocity of $\vsys=\qtyvcent$, in the diffuse \Hi to the dense \Hi.
    (f)~The \los velocity difference between the dense and diffuse \Hi.
  }
  \label{fig:decompose_result}
\end{figure*}

We report in \autoref{fig:decompose_result} the results of our double-Gaussian decomposition.
The dense and diffuse \Hi (panels a and~b) divides roughly at a column density of \qtydiffusedivision.
As expected from the decomposition criteria, the spatial distribution of dense \Hi is similar to the interferometric observations (\autoref{app:ssec:mom0_comp}), and its velocity field (panels c and~g) resembles that of the total \Hi (\autoref{fig:deconv_result}b).
Both moment-1 maps have an inflow pattern that the zero-relative-velocity line and the minor axis are misaligned,%
\footnote{
  The misalignment is due to the $\sin\phi$ projection of inflow velocities, where $\phi$ denotes the azimuthal angle from the major axis.
}
while the dense-associated diffuse \Hi does not have a similar misaligned pattern (panel~h, within the gray contour).

\citetalias{2024ApJ...968...48W} found that the diffuse \Hi in the inner region rotates at a smaller velocity than the dense one does, and the velocity difference shown in \autoref{fig:decompose_result}(f) exhibits this ``lagging'' pattern at a better resolution.
There are some fine structures on the velocity difference map at the east or west sides of the dense-\Hi disk, possibly relating to tidal stripping and shocks (also see \autoref{ssec:rp_shock}).
Nonetheless, the overall behavior of the dense-associated diffuse \Hi is corotating with the dense \Hi, while lagging behind it with a median velocity offset of \qty{12}{\km\per\s} (\autoref{fig:decompose_result}f), or a relative extent of \qty{23}{\percent} (\autoref{fig:decompose_result}e).

As discussed in \citetalias{2024ApJ...968...48W}, the lag in velocity often indicates extraplanar gas accretion through the galactic fountain mechanism \citep{2017ASSL..430..323F}.
It was also found that the denser gas tends to be more responsive to spiral arms and thus to form radial inflows \citep{2014ApJ...784....4C}.
These patterns confirm the complex but typical physical processes happening close to the star-forming disk.
In the following, we more focus on the low-column density \Hi extending radially far into the CGM beyond the star-forming disks.
This outlying \Hi extends the radial range for the disk radial inflow to happen, and serves as a useful test of \Hi--CGM interaction.

\subsection{The Position--Velocity Distribution of Tidal Structures and the Possible Merger History}
\label{ssec:tail}
\begin{figure*}
  \centering
  \includegraphics{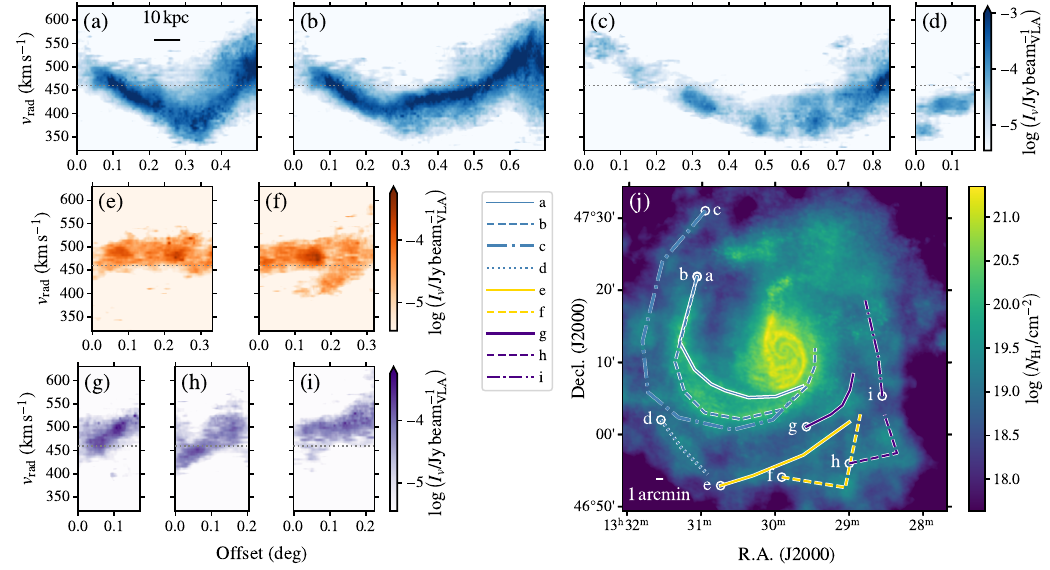}
  \caption{
    Identified structures in M51 (j) and their corresponding PV diagrams (a--i).
    The structures are divided into three groups and identified with different colors: a--d (blue), e--f (orange), and g--i (violet).
    The redshift of M51a center ($\vsys=\qtyvcent$) is plotted on the PV diagrams as gray dotted lines.
    In panel~(j), white circles mark the line end corresponding to the left side of PV diagrams.
  }
  \label{fig:structure_pv}
\end{figure*}

To infer the possible origins of the large-scale \Hi structure, we plot the position--velocity (PV) diagrams in \autoref{fig:structure_pv}(a)--(i) along the \ang{;1;}-wide section lines shown in \autoref{fig:structure_pv}(j).
The labeled white circles in panel (j) mark the left-hand edge of the corresponding PV diagrams in panels (a)--(i).

We identify three ridges along the tidal tail (a--c).
Their PV diagrams indicate that the tail is a single continuous structure stemming from the southwestern edge of the dense-\Hi disk.
The asymmetric wings in panels (a) and~(b) in the velocity direction imply the lagging or leading of diffuse \Hi by tidal and/or ram-pressure forces.
The structure (d) has a similar PV shape and may have been broken from the major tidal tail.

The \Hi features traced by lines (e)--(i) are more complicated in their morphology.
Along lines (e) and~(f), the velocity is almost constant at \qty{\sim475}{\km\per\s}, slightly redshifted relative to the M51a center ($\vsys=\qtyvcent$).
The blueshifted feature on the right side of panel~(f) is from the projection of a different structure that is also traced by line~(h).
Panels (g) and~(i) show a positive velocity gradient toward the north, where their lines recede and connect with the inner disk.

In summary, from the inner \Hi disk protrude out two relatively coherent groups of tail-like structures (a--d and e--i).
Those in the former are elongated and linked, while those in the latter are intermittent and segmented.
These two groups become kinematically similar when reaching the dense disk in the west.
It is possible that at the beginning, these two groups were produced together during one major encounter between the two galaxies, but then the second has been further disrupted by another major encounter, as indicated by previous simulations (see \autoref{sssec:comp_simu}).

\subsubsection{The Comparison with the Hydrodynamic Simulation of \texorpdfstring{\citet{2010MNRAS.403..625D}}{C. L. Dobbs et al. (2010)}}
\label{sssec:comp_simu}
\citet{2010MNRAS.403..625D} made a hydrodynamical simulation of M51 based on the orbital initial conditions from \citet{2003Ap&SS.284..495T}.
They largely recovered the \Hi structure detected by \citet{1990AJ....100..387R} with VLA, which was similar to the THINGS image (shown in \autoref{fig:mom0_comp}b of \autoref{app:ssec:mom0_comp}), but they missed the radially extended structure in our jointly deconvolved image of \autoref{fig:deconv_result}(a).
The simulation did not include a CGM halo and therefore any hydrodynamic CGM--ISM interaction.
Thus, it can help to distinguish the roles of tidal interaction with M51b (when the simulation successfully reproduced structures) and hydrodynamic effects of \Hi--CGM interactions (when the simulation failed to reproduce structures) in forming the observed large-scale \Hi structure.

In their simulation, M51a formed two large tidal tails as M51b approached from the north \qty{\sim300}{\Myr} ago \citep[see the top middle panel of Figure~4 in][]{2010MNRAS.403..625D}.
One of them later became the tail currently observed to the east of M51a (structures a--d).
However, there are two significant differences from the observation, which possibly come from the absence of the interaction with CGM:
(1)~their simulated tidal tail opens more widely, and (2)~the simulated system lacks a diffuse envelope that should have extended to \qty{\sim50}{\kpc}.

In the simulation, \qty{\sim120}{\Myr} ago, M51b pierced through and thus disrupted the other tail when both M51b and the tail orbited to the south (the top right panel of their Figure~4).
Since then, a great portion of the shattered tail was dragged to the north, but a narrow stripe extending to the west also formed in the simulation.
This predicted structure roughly corresponds to but is very different from the broad, intermittent, and segmented structures labeled as (e)--(i) in \autoref{fig:structure_pv}(j), which extend to the south and then bend eastward.
Especially, structures (e) and~(f) have no obvious velocity gradient, implying that the radial velocities have previously decreased, possibly due to the drag of ram pressure or newly accreted CGM gas, both of which were not considered by \citet{2010MNRAS.403..625D}.

In summary, the encounter simulated in \citet{2010MNRAS.403..625D} was largely consistent with the dense-\Hi structures, but missed most of the new diffuse-\Hi features shown in FEASTS data, possibly because they did not include the CGM--ISM interaction, which is the major theme of this study.
In the simulation, the interaction started \qty{\sim300}{\Myr} ago, and the second encounter happened \qty{\sim120}{\Myr} ago.
These results set a rough time range for the diffuse \Hi to grow and survive.

\subsection{The Relation between \Hi-gas Column Densities and Velocity Dispersions}
\label{ssec:localized}
From now on, we focus on the statistical behavior of localized properties, such as the column density \NHi and the velocity dispersion \sigv.
We use the previously defined \los sample (with $\sigv<\rnghighlydisperse$) in the following analyses, and exclude the stellar disk (i.e., $r<\Rzs$).

\begin{figure*}
  \centering
  \includegraphics{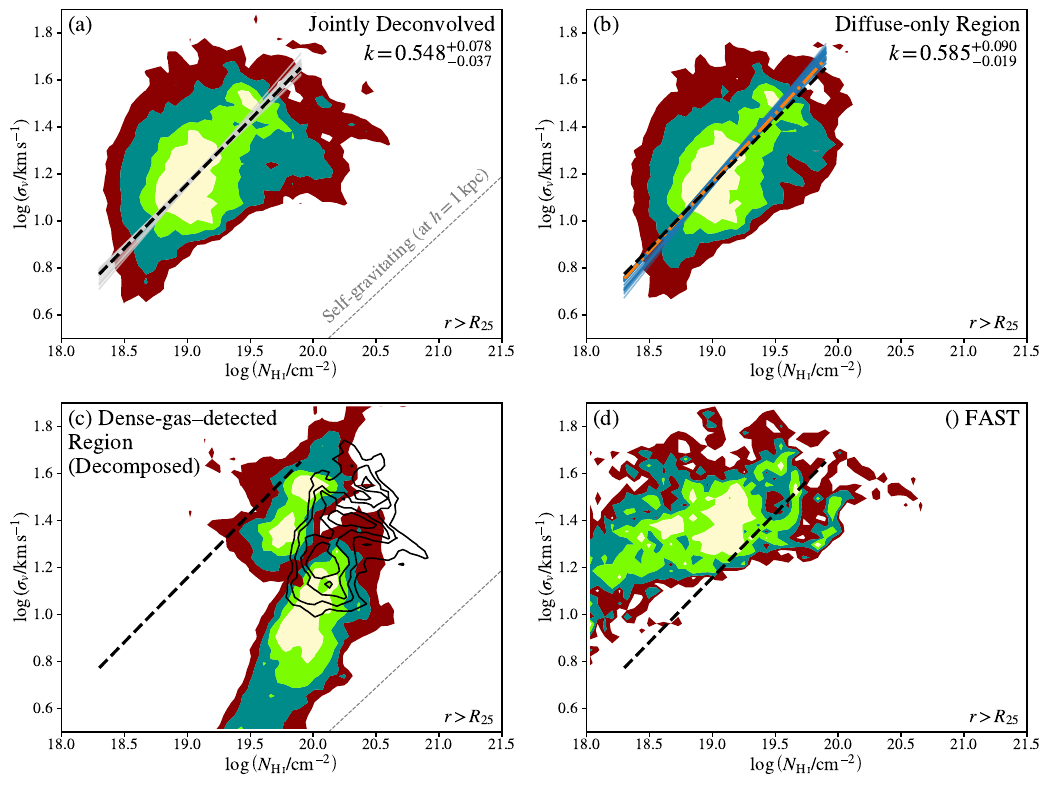}
  \caption{
    The pixelwise distribution of \Hi column density \NHi and the moment-2 map value of \sigv of (a)~the jointly deconvolved cube, (b)~diffuse-only \Hi, (c)~dense and dense-associated diffuse \Hi, and (d)~the FAST cube, given as the filled colored contours.
    In panel~(c), the dense and dense-associated diffuse \Hi corresponds to the lower and upper branches, respectively, and the distribution before decomposition is given as black contours.
    In all panels, the contours enclose \qtylist{90;75;50;25}{\percent} of the corresponding data from the outside in.
    We conduct bisector least-squares linear fitting between the \sigv and \NHi for pixels within the \qty{75}{\percent} contour in panels (a) and~(b).
    The fitted line of panel (a) and~(b) are plotted as the black dashed line (repeated in all panels) and orange dashed--dotted line, respectively.
    The slopes are roughly \num{0.5}, reported at the corner of each panel.
    The uncertainties of slopes are the variation range when the threshold contour changes from \qty{90}{\percent} to \qty{50}{\percent} (the gray and blue solid lines in two panels).
    In panels (a) and~(c), we plot as the gray dashed line the anticipated \sigv--\NHi relation if the turbulence pressure is balanced by the self-gravity of gas.
    The gas thickness is assumed to be \qty{1}{\kpc} as an upper limit (see the thickness estimation in \autoref{ssec:radial_property});
    the line would move down with a smaller thickness.
  }
  \label{fig:N_sig}
\end{figure*}

A change in $\NHi=\nHi h/\cos i$ might come from the variation in either \Hi number density (\nHi) or thickness ($h$), both of which are related to the local turbulence level or velocity dispersion.
In \autoref{fig:N_sig}(a), we plot the pixelwise distribution of \NHi and \sigv of the jointly deconvolved cube with contours enclosing \qtylist{90;75;50;25}{\percent} of the \los sample.
Here, \sigv is represented by the value of moment-2 map.
The distribution for diffuse-only \Hi is given in \autoref{fig:N_sig}(b).

Both distributions in panels (a) and~(b) show a positive relation between \sigv and \NHi when $\NHi\lesssim\qty{1e20}{\per\cm\squared}$.
The upper envelopes of the distributions in \autoref{fig:N_sig}(a) and~(b) are less sharp than the lower envelopes, possibly because of the projection of structures along the same \los, leading to the overestimation of \sigv at a given \NHi.
This trend is driven by the behavior of diffuse-only \Hi because the dense \Hi is detected only when the total $\NHi>\qtydiffusedivision$.

We use bisector least-squares linear fitting to evaluate the ridge lines of the distributions in \autoref{fig:N_sig}(a) and~(b), using the relatively confident \los{}s within the \qty{75}{\percent} contour.
We plot the result as the black dashed line and orange dashed--dotted line in panels (a) and~(b), respectively, with a slope of \numkcomb and \numkfaint.
This final fitting result is the bisector line of the following two fittings:
(1)~we fit \sigv against \NHi with the weighting of $1/n$, where $n$ is the number of \los{}s in each $\log\NHi$ bin, and
(2)~we fit \NHi against \sigv, weighted by the inverse of the \los number in $\log\sigv$ bins.
The systematic uncertainties of slopes are estimated as the range of variations with the threshold contour level changing from \qty{90}{\percent} to \qty{50}{\percent} (gray and blue solid lines in two panels).
We confirm with mock tests that the random uncertainties from the individual values of \NHi and \sigv do not contribute significantly (\num{<0.020}) to the slope uncertainty.
For reference, the bisector-fitted line in panel~(a) is repeated in all four panels.

This slope of \num{\sim0.5} coincides with that of self-gravitating gas like the dense molecular clouds \citep[e.g.,][]{2009ApJ...699.1092H,2015ApJ...801...25L,2018ApJ...860..172S,2020ApJ...901L...8S}, but the \NHi expected for self-gravitating at a given \sigv should be much higher (the gray dashed line in panels a and~c), inconsistent with the actual distribution.
The \Hi in this regime is mostly located outside the optical disk, and should be strongly regulated by CGM--\Hi interactions.

The distribution in panel~(a) flattens at $\NHi\gtrsim\qty{1e20}{\per\cm\squared}$, where the dense \Hi is more dominant in mass.
In this regime, processes originating in the stellar disks, e.g., the gravity and stellar (and possibly active galactic nuclei) feedback, should play a more important role than the CGM--\Hi interaction in regulating the relation \citep[e.g.,][]{2018MNRAS.477.2716K}.
A deeper investigation of this regime is beyond the scope of this paper.

After we decompose the dense and diffuse components (\autoref{fig:N_sig}c) in the dense-\Hi--detected region, the \sigv--\NHi distribution clearly separates into two branches.
The diffuse branch (the upper one) roughly follows the same trend of the positive correlation between \sigv and \NHi.
However, the diffuse branch effectively consists of the residuals of the Gaussian fitting for dense \Hi, therefore having a low S/N for a confident \sigv measurement.
We thus refrain from making further interpretations.

In \autoref{fig:N_sig}(d), the trend of FAST data is shallower compared with the jointly deconvolved one (black dashed line), suggesting the influence of beam-smoothing effects.
Therefore, the joint deconvolution is essential for recovering the \Hi distribution and revealing its correct properties.

\subsection{The Radial Distribution of Number Density and Thickness of \Hi}
\label{ssec:radial_property}
\begin{figure*}
  \centering
  \includegraphics{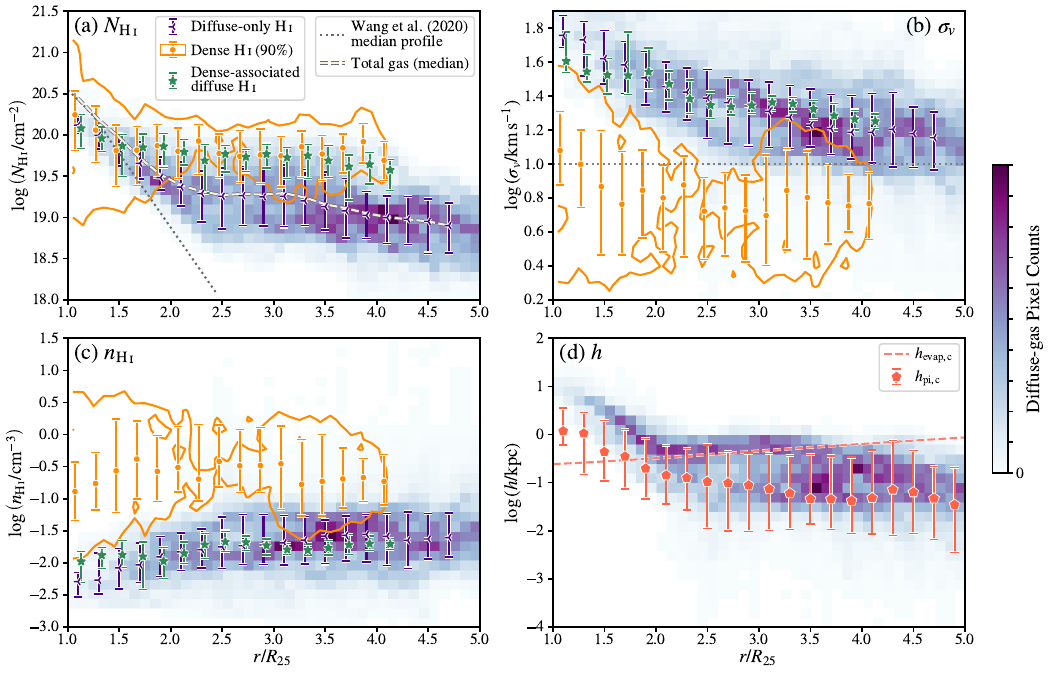}
  \caption{
    The radial profiles of the following \Hi properties: (a)~column density \NHi, (b)~velocity dispersion \sigv, (c)~volumetric density \nHi, and (d)~thickness $h$.
    The radius $r$ from the center is normalized by optical $\Rzs=\qtyRzs$, after considering the inclination.
    The diffuse-only \Hi is given as the violet shading, and the ``Y''-shaped symbols and error bars are the median and 16th and 84th percentiles in each $0.2\Rzs$-wide bin.
    Similarly, the orange dots and filled green stars correspond to dense \Hi and dense-associated diffuse \Hi, respectively.
    The orange contour encloses \qty{90}{\percent} of the dense \Hi pixels.
    In panel~(a), the median profile of total \Hi is the white dashed line.
    The statistical median \Hi profile from \citet{2020ApJ...890...63W} is the gray dotted line, linearly extrapolated to $2\RHi$;
    the \Hi radius \RHi is defined at an \Hi surface density of \qty{1}{\Msun\per\pc\squared}.
    In panel~(b), the horizontal gray dotted line is \qty{10}{\km\per\s} for reference.
    In panel~(d), the minimal thicknesses to survive thermal evaporation (\hevapc) and photoionization (\hpic) are given as the red dashed line and the filled red pentagons with error bars, respectively.
  }
  \label{fig:n_h}
\end{figure*}

The radial distribution of column density \NHi is shown in \autoref{fig:n_h}(a) for different \Hi components of the \los sample, after we exclude the stellar disk ($r<\Rzs$).
The \NHi profiles are largely flat at the \qty{\sim1e19}{\per\cm\squared} level beyond $2\Rzs$, echoing our earlier point that this system is an interesting laboratory to study the \Hi--CGM interaction.
In comparison, the slope of the M51 \NHi profile is much shallower than the median radial profile (gray dotted line in panel~a) of \Hi in galaxies based on interferometry data \citep{2020ApJ...890...63W}, the profiles of nine other galaxies that have both THINGS and FEASTS data (see \autoref{app:sec:FEASTS_NHi}, data from \citetalias{2024ApJ...973...15W}), and most of the galaxies observed by FEASTS \citep{2025ApJ...980...25W}.

In \autoref{fig:n_h}(b), the velocity dispersion \sigv of diffuse \Hi decreases with radius.
The trend seems like the extension of what has been observed in the $\NHi>\qty{1e20}{\per\cm\squared}$ regime \citep[e.g.,][]{2019A&A...622A..64B}.
A non-negligible fraction (\qty{20}{\percent}) of \los{}s at $r>3\Rzs$ have $\sigv<\qty{10}{\km\per\s}$ (gray dotted line in panel~b), implying that the diffuse \Hi with so low a column density is not very efficiently heated by the CGM and UV background.
We will discuss this further in \autoref{sssec:hi_survive}.

Assuming that the \Hi turbulence pressure $1.4\mprot\nHi\sigv^2$ is in equilibrium with the sum of CGM pressure and midplane \Hi gas weight,%
\footnote{
  We calculate the \Hi gas weight felt by the diffuse \Hi as $\pi G\Sigdiffuse\times(\Sigdiffuse/2+\Sigdense)$, i.e., its self-gravity plus the gravity from embedded dense \Hi \citep[see Equation~7 of][]{2010ApJ...721..975O}.
  The weight felt by dense \Hi is calculated as $\pi G(\Sigdiffuse+\Sigdense)^2/2$, i.e., the total \Hi gas weight, because the dense \Hi is much thinner than the diffuse one;
  a demonstrative diagram can be found in \citet[Figure~A.1]{2020ApJ...892..148S}.
  Here, $G$ is the gravitational constant, and \Sigdiffuse (or \Sigdense) denotes the deprojected diffuse (dense) \Hi surface density including the contribution of helium.
}
we calculated per pixel the volume density \nHi and the thickness $h=\cos(i)\NHi/\nHi$ for different \Hi components at $r>\Rzs$.
Although the whole galaxy is a dynamically unstable interacting system, the localized gas tends to be in thermodynamic equilibrium.
Any localized inequilibrium should dissipate within a sound-crossing time of \qty{\sim10}{\Myr} across the thickness, much shorter than the merging timescale.
Here, \mprot is the mass of proton, and the local inclination angle $i$ is taken from the tilted-ring modeling (\autoref{ssec:vfieldmod} and \autoref{app:sec:vfieldmap}).
The CGM pressure, calculated as $\pcgm = \rhocgm k\tsb{B}\Tcgm/\mucgm$, is three times higher than the \Hi gas weight
within $1.5\Rzs$ and has a larger contribution beyond.
We take the average particle weight $\mucgm=0.6\mprot$.
The virial temperature ($\Tcgm=\qty{1.06e6}{\K}$) and gas density profile (\rhocgm) are calculated from the $M\tsb{vir}$, assuming a typical $\beta$ model following Appendices D and~E of \citetalias{2023ApJ...944..102W}.%
\footnote{
  The virial mass $M\tsb{vir}$ is assumed to have a critical overdensity of \num{101}.
  We estimate the total CGM gas mass using Equation~(5) of \citet{2015MNRAS.446.2629E}.
  The density profile has $\beta=0.64$ and a core radius of $0.15\Rsoo$, where \Rsoo is the radius within which the average density is \num{500} times the critical density.
}
We ignore the insignificant contribution (\qty{<10}{\percent}) of \Hi thermal pressure.
We assume that the cool gas is mostly neutral, which is validated a~posteriori in \autoref{sssec:hi_survive}.
The magnetic pressure is not considered because it is not well constrained observationally \citep[e.g.,][]{2015ApJ...799...35V,2017ARA&A..55..111H,2021ApJ...921..128B,2023ApJ...944...86M}, implying an unknown level of over- and underestimation of \nHi and $h$, respectively.

The results are shown in \autoref{fig:n_h}(c) and~(d).
Both dense and diffuse \Hi have an almost constant volumetric density, especially out of $2.5\Rzs$.
The diffuse \nHi fluctuates around \qty{2e-2}{\cm^{-3}}, and the dense \nHi is \qty{\sim1}{\dex} higher, consistent with the theoretical values typical of their respective column densities \citep{2013MNRAS.430.2427R}.
The constant \nHi implies a constant cooling rate of diffuse \Hi, possibly serving as the cooling interface between hot CGM and dense \Hi \citepalias[also see][based on coarser measurements]{2024ApJ...968...48W}.

The \Hi thickness $h$ decreases with radius, contrary to the \Hi flaring phenomenon previously found at smaller radii based on a static hydrodynamic modeling of the balance between the gas turbulence pressure and the off-plane gravity \citep[e.g.,][]{2019A&A...622A..64B}.
For a flaring disk, the $h$ increases with $r$ due to the increasingly weaker gravitational confinement.
From our calculation, the underlying reasons for this contrary trend are the relatively high CGM pressure and the low \Hi velocity dispersion.
The CGM takes over the confining role, pressing the \Hi to rather high volume densities, and the low velocity dispersion of \Hi implies the existence of efficient kinetic energy dissipation or cooling (\autoref{ssec:tml_scenario}).

\subsubsection{The Critical Thicknesses of Photoionization and Thermal Evaporation}
\label{sssec:hi_survive}
There is the possibility that the diffuse \Hi surrounding M51 could be destroyed by the CGM through thermal conduction or photoionization.
Meanwhile, the existence of these \Hi structures indicates that the net heating is not that efficient in causing a change in the phase distribution.
In this section, we estimate criteria for these two possible mechanisms to take effect.

\paragraph{Photoionization.}
We estimate the condition for significant photoionization following \citetalias{2023ApJ...944..102W}, using their \emph{Cloudy} simulation results \citep{1998PASP..110..761F}.
The simulation assumes that the gas is smooth;
therefore, any derived critical condition for surviving photoionization should be loosened if the actual gas has clumpy structures.
Given that the leakage of UV photons along the spiral disk plane is weak, we assume that the ionizing photons in the outer disk (${>}\Rzs$) mainly come from the cosmic background radiation, described by the ionization parameter $U$ as $\qty{1e-6}{\cm^{-3}}/\nHi$ \citep{2011ApJ...733..111T} with the default Cloudy background setting.

\begin{figure}
  \centering
  \includegraphics{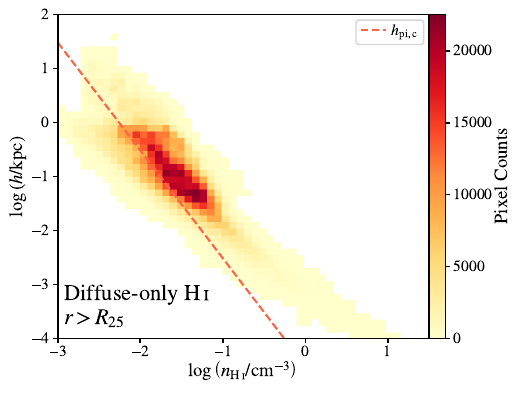}
  \caption{
    The distribution of diffuse-only \los{}s in the \Hi thickness $h$--volumetric density \nHi space, after we exclude the stellar disk and high-\sigv regions.
    The critical thickness \hpic for a neutral ratio of \qty{50}{\percent} due to photoionization is plotted as the red dashed line.
    In total, \qty{78}{\percent} of \los{}s and \qty{91}{\percent} of diffuse-only \Hi mass are self-shielded from photoionization.
  }
  \label{fig:n_h_pi}
\end{figure}

In \autoref{fig:n_h_pi}, we plot the critical thickness for a neutral ratio of \qty{50}{\percent} (\hpic, red dashed line) over the $h$--\nHi distribution.
With the stellar disk and high-\sigv regions excluded, \qty{78}{\percent} of \los{}s and \qty{91}{\percent} of diffuse-only \Hi mass are self-shielded from photoionization.
The diffuse \Hi is reasonable to be treated as neutral when we calculate the pressure balance in \autoref{ssec:radial_property}.
In \autoref{fig:n_h}(d), we also provide the median \hpic within bins of width $0.2\Rzs$, which roughly follows the lower envelope of the majority of $h$ values.

\paragraph{Thermal Evaporation.}
We take the \citet{1977ApJ...211..135C} model to calculate the minimal survival thickness (\hevapc) that makes the saturation parameter $\sigma_0=(\Tcgm/\qty{1.5e7}{\K})^2/(\nHcgm\hevapc/\unit{\per\cubic\cm\pc})$ reach the critical value of \num{0.027} \citep{1977ApJ...215..213M}, above which there would be evaporation flow.

The \hevapc profile is plotted in \autoref{fig:n_h}(d) as the red dashed line.
Beyond $2.5\Rzs$, most of the \Hi gas is possibly susceptible to evaporation.
Following \citet{1977ApJ...211..135C}, we estimate that at \qty{38}{\percent} of the diffuse-only \los{}s, the \Hi should evaporate in less than \qty{\sim120}{\Myr}, the time since the last encounter between M51a and M51b (\autoref{sssec:comp_simu}).

In this evaporation model, the \Hi is assumed to exist as individual spherical cloudlets with a size of ${\sim}h$;
if the \Hi is a slab, the evaporation could be much less effective \citep{1977Natur.266..501C}.
However, the turbulent nature of cool gas in the CGM tends to rule out the slab morphology \citep{2023ApJ...955L..25C}.
Another possible caveat is that the magnetic field could significantly suppress the conduction \citep{1962pfig.book.....S,1965RvPP....1..205B,2021MNRAS.502.1263K} if its orientation mainly follows the surface of \Hi gas.
However, the latest observations suggest that magnetic lines loop perpendicular to the \Hi disk plane \citep{2019A&A...632A..13S}.
If we use calibrated equations of criteria from the latest magnetohydrodynamics simulations \citep[e.g., Figure~5 of][]{2020MNRAS.492.1841L}, the obtained critical thickness is very similar.

One possibility is that strong cooling processes are at work, and are able to compensate for the thermal conduction loss and sustain the extended diffuse \Hi.
We propose a scenario of turbulent-mixing gas cooling and accretion in \autoref{sec:tml_cool}.

\section{Turbulent-mixing Gas Accretion}
\label{sec:tml_cool}
\begin{figure*}
  \centering
  \includegraphics{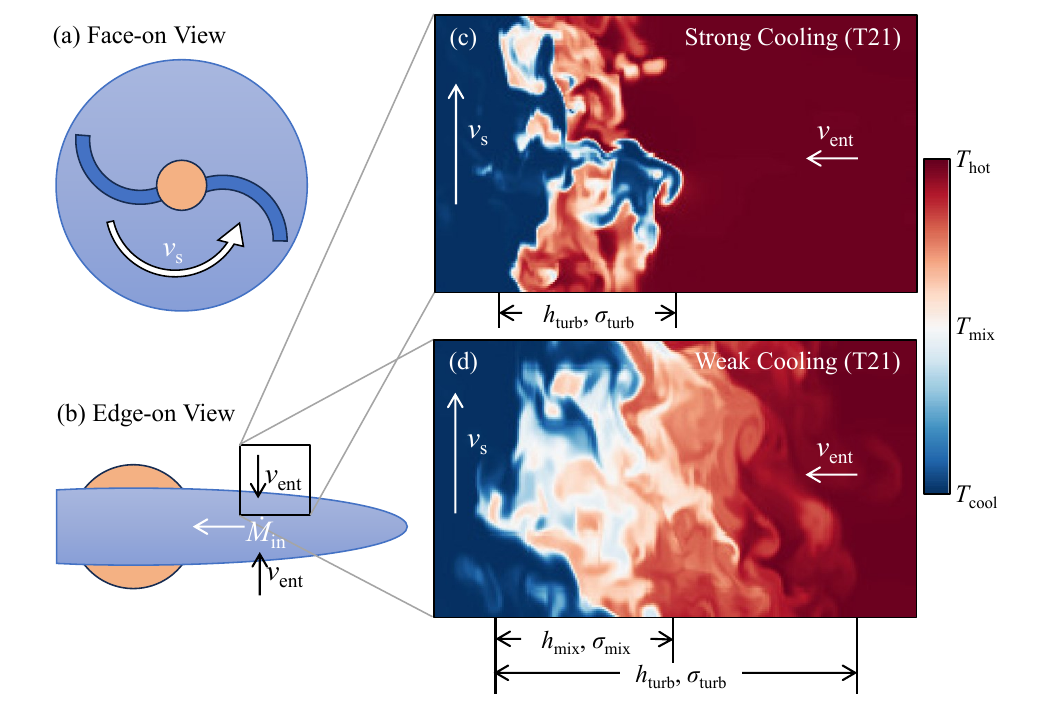}
  \caption{
    The schematic plots of the turbulent-mixing gas cooling and accretion scenario.
    (a)~The face-on view of the cool \Hi disk rotating at a velocity of \vs.
    (b)~The edge-on view.
    The shearing between the \Hi disk and the CGM forms TML, where the mixed gas at $\Tmix\sim\qty{1e5}{\K}$ efficiently cools, and is entrained into the \Hi disk at a velocity of \vent perpendicular to the disk.
    The gas condensed from the CGM has lower angular momentum, and would flow in.
    The inflow rate is denoted as \dMin.
    (c)~A slice of the temperature field in the strong-cooling regime (see the main text in \autoref{ssec:tml_scenario}).
    The direction of \vs and \vent are shown by the white arrows.
    The TML thickness \hturb is bracketed by black arrows, and \sigturb is the velocity dispersion within the layer.
    (d)~The same as panel~(c), but for the weak-cooling regime.
    The well-mixed region is labeled with thickness \hmix and dispersion \sigmix.
    The temperature slices in panels (c) and~(d) are taken from Figure~7 of \citetalias{2021MNRAS.502.3179T}.
  }
  \label{fig:tml_cartoon}
\end{figure*}

In this section, we propose a scenario whereby the ISM--CGM interaction is governed by a turbulent mixing layer \citep[TML;][]{1990MNRAS.244P..26B}.
The TML theory highlights the efficient radiative cooling of the intermediate-temperature (\qty{\sim1e5}{\K}) gas in TML, which is generated from the mixing at the shearing interface between the cool- and hot-phase gas.
In our proposed scenario, the \Hi not only efficiently condenses from the CGM, but also flows inward to fuel the star formation.
Globally, these two processes correspond to the gas accretion onto \Hi disk and into the star-forming optical disk.
This process may help the diffuse \Hi to survive thermal conduction, and to replenish the galactic \Hi reservoir.

We emphasize that this is just one possible scenario, but it has largely been overlooked in the literature and is thus interesting to explore.
It is worth pointing out that the TML is a physical component generally unresolved in the existing cosmological simulations, and is difficult to directly observe in observations \citep{2023ARA&A..61..131F}.
Our study assumes a simple model, and obtains a possible ensemble-averaged view of the phenomenon.
Specifically, tidal interaction in the M51 system dragged out its large-scale \Hi structure, which serves as a seed for future cooling due to shearing in TMLs.
Keeping in mind that other physical processes may be at play to influence the results, we nonetheless obtain interesting constraints, which may hopefully inspire future efforts with multiwavelength observations and theoretical recipes.

\subsection{The Scenario}
\label{ssec:tml_scenario}
This paper aims to provide a feasibility test for the turbulent-mixing gas cooling and accretion on galaxy scale, rather than an observational verification of the scaling relations from the numerous 3D idealized simulations of TML \citetext{e.g., \citealp{2016MNRAS.462.4157A,2018MNRAS.477.3293P}; \citealp[hereafter \citetalias{2019MNRAS.487..737J}]{2019MNRAS.487..737J}; \citealp{2019MNRAS.484.1100M,2020MNRAS.494.2641M,2020ApJ...894L..24F,2020MNRAS.492.1970G}; \citealp[hereafter \citetalias{2021MNRAS.502.3179T}]{2021MNRAS.502.3179T}; \citealp{2023MNRAS.520.2148Y,2024MNRAS.532.2965A}}.
These simulations have explored in different geometric configurations the relations between the kinematic and geometric properties of TML, and the rate of hot gas cooling and entrainment.
A conclusive census on the exact scaling relations has still not been reached, but these studies share similar physical insights, and give similar numerical predictions for the TML properties.

We summarize the key points of our scenario as the cartoons in \autoref{fig:tml_cartoon}.
We consider the shearing (at speed \vs) between the cool \Hi disk ($T\tsb{cool}\lesssim\qty{1e4}{\K}$) and the hot and rather static CGM ($T\tsb{hot}\gtrsim\qty{1e6}{\K}$) from the rotation of the former (\autoref{fig:tml_cartoon}a).
Due to the Kelvin--Helmholtz instability, TML forms between them with a temperature of $\Tmix\sim\sqrt{T\tsb{cool}T\tsb{hot}}\sim\qty{1e5}{\K}$ (\autoref{fig:tml_cartoon}c and~d).
From the equation of momentum conservation, the thickness \hturb and \los velocity dispersion (in the direction of \hturb) \sigturb of TML are related as \citetext{\citealp{2004ApJ...616..669K,2016MNRAS.455.2042W}; \citetalias{2019MNRAS.487..737J}}
\begin{equation}\label{eq:h_sig}
  \frac{\hturb}{L}\sim\frac{\sigturb^2}{\vs^2}.
\end{equation}
Here, $L$ is the typical scale of TML along the shearing interface.

When \vs is subsonic within the CGM%
\footnote{
  The speed of sound in M51 CGM is \qtycscgm, larger than most of the shearing velocity (see \autoref{ssec:vfieldmod}).
}
\citep{2023MNRAS.520.2148Y}, the strong cooling at \Tmix causes the well-mixed gas in TML to condense into gas of $T\tsb{cool}$ in the \Hi phase, which we directly observe in this work.
After merging with the cold layer, the condensed \Hi would flow in because of their lower angular momentum inherited from the CGM\@.
In a dynamical equilibrium, the cooling and condensation rates are expected to be comparable to the radial inflow rate.
We expect that the condensed \Hi replenishes and makes up a significant part of the diffuse \Hi, and their properties, such as thickness and dispersion, may correlate with those of TML\@.

\citetalias{2021MNRAS.502.3179T} categorized the cooling process within TMLs into the weak and strong regimes, according to the relative importance of radiative cooling and turbulence, which is characterized as the ratio of their timescales, i.e., the Damk\"ohler number $\Da=\tturb/\tcool$.%
\footnote{
  \citetalias{2021MNRAS.502.3179T} estimated \tturb as the eddy turnover time ($\Leddy/u'$), where \Leddy is the outer scale of eddies, and $u'\approx\sqrt{3}\sigv$ is the 3D turbulent velocity.
}
The TMLs in these two regimes have distinct physical structures, and their entrainment rates of hot gas show different scaling relations.
In the following, we briefly summarize features of each regime, and discuss in each case properties of possible \Hi observables, such as the column density \NHi and \los velocity dispersion \sigv.

\paragraph{The Strong Cooling ($\Da>1$)}
In this regime, the cooling is efficient compared with the turbulent diffusion, resulting in a thin but highly corrugated mixing and cooling interface, which separates the TML into distinct cool- and hot-phase regions.
Most of the cooling processes take place in the the thin wrinkled interface (\autoref{fig:tml_cartoon}c), of which the increased interface area helps the global cooling rate to increase.
The cool gas occupies roughly half of the strong-cooling TML volume \citepalias{2021MNRAS.502.3179T}, and shares the turbulence level at the TML scale.

In this case, the \Hi that we observe traces the cool-phase region of the TML, approximately satisfying
\begin{align}
  \NHi  & = 2\nHi[cool]\times\frac{1}{2}\hturb,
  \label{eq:NHi:strong_cool}                    \\
  \sigv & = \sigturb,\label{eq:sig:strong_cool}
\end{align}
when the TML accounts for a significant part of the \Hi.
The factor \num{2} accounts for both sides of the shearing, and $1/2$ is the approximate filling factor of cool gas in the TML \citepalias[also see Figure~6 of][]{2021MNRAS.502.3179T}.
Here, the measured \NHi and \sigv are the averaged value within the \qtyphysVLAfwhm beam.

The \Hi in the TML is in direct pressure equilibrium with the hot gas in and outside the TML\@.
Their pressures are mainly contributed by turbulence and thermal pressure, respectively, i.e.,
\begin{equation}\label{eq:pressure_strong}
  \pcgm = \rhoHi[cool]\sigturb^2 = 1.4\mprot\nHi[cool]\sigturb^2.
\end{equation}

\paragraph{The Weak Cooling ($\Da<1$)}
In this regime, the cooling is inefficient, making the TML thicker than the cooling interface, which is on the cold-gas side of the TML (\autoref{fig:tml_cartoon}d).
It is in this thin interface of thickness \hmix that the hot and cold gas fully mixes into gas of temperature \Tmix and significantly cools.
Across the whole TML of $\hturb>\hmix$, the temperature smoothly transits from $T\tsb{hot}$ to $T\tsb{cool}$, similar to the case within a thermal conduction front.
The velocity dispersion within the cooling interface \sigmix is smaller than the global value \sigturb within the whole TML \citepalias[Figure~6 of][]{2021MNRAS.502.3179T}.

The well-mixed gas in the cooling interface condenses into numerous \Hi cloudlets, of which each has a mass $\mcloud=1.4\mprot\CHi$, where \CHi denotes the number of neutral hydrogen atoms in one cloudlet.
We treat the cloudlets as individual particles, each having its own internal thermal temperature and velocity turbulence, while they together form a mist with a cloud number density \ncloud and rms velocity dispersion \sigcloud.
As the ensemble measure within a beam-sized aperture, \sigcloud is a robust proxy for the turbulence of the mixed gas before condensation, \sigmix \citep{2018ApJ...854..167G}.

In this regime, the observed \Hi is the condensed cloudlets, and the measurements can be written as
\begin{align}
  \NHi  & = 2\nHi\hmix = 2\ncloud\CHi\hmix < 2\ncloud\CHi\hturb,
  \label{eq:NHi:weak_cool}                                       \\
  \sigv & = \sigmix < \sigturb.\label{eq:sig:weak_cool}
\end{align}
The pressure in cooling layer is dominated by the cloudlets, which fill roughly half the volume and account for most of the mass.
Similar to the \autoeqref{eq:pressure_strong}, the pressure equilibrium can be written as
\begin{equation}\label{eq:pressure_weak}
  \pcgm = \rhocloud\sigmix^2,
\end{equation}
where the cloudlet density averaged over the whole cooling interface is $\rhocloud=1.4\mprot\ncloud\CHi$.

\subsection{Expectations for the Observations}
\label{ssec:tml_expectation}
Considering the scenario introduced above, we expect the following observational results in a statistical sense.
\begin{enumerate}
  \item \emph{A relation of $\NHi\sim\sigv^2$ with scatter.}
        If most of the TMLs are in the strong-cooling regime, we expect that $\NHi\propto\hturb\sim\sigturb^2=\sigv^2$ (Equations \ref{eq:h_sig}--\ref{eq:sig:strong_cool}).
        Such a relation has been found in \autoref{ssec:localized} (\autoref{fig:N_sig}).
        If some of the TMLs are in the weak-cooling regime, data points would deviate from the relation, as $\hmix<\hturb$ and $\sigmix<\sigturb$.
  \item \emph{The thickness of the \Hi layer \hHi.}
        The pressure equilibrium relations of Equations \eqref{eq:pressure_strong} and~\eqref{eq:pressure_weak} indicate that an effective \Hi thickness \hHi can be derived as the $h$ in \autoref{ssec:radial_property}, but the interpretation is different.

        In the strong-cooling regime, \hHi is half the TML thickness \hturb (\autoref{eq:NHi:strong_cool}).
        Considering that the \Hi structures have two sides of shearing interface, we could estimate that $\hturb\lesssim 2\hHi = h$, which is roughly \qtyrange{47}{290}{\pc} (25th and 75th percentiles out of $2\Rzs$), consistent with the typical values of \qty{\sim100}{\pc} in simulations \citetext{e.g., \citetalias{2019MNRAS.487..737J}; \citetalias{2021MNRAS.502.3179T}; \citealp{2023MNRAS.520.2148Y}}.

        In the weak-cooling regime, \hHi is roughly the cooling interface thickness $\hmix\approx h/2$ (\autoref{eq:NHi:weak_cool}).
        In this case, the \Hi is highly clumpy, and is thus more capable of surviving photoionization.
  \item \emph{The possibility of estimating the CGM gas entrainment velocity (\vent).}
        The corresponding cooling rate per surface area can be written as $\dd\dMcool\big/\dd A=\rhocgm\vent$ (more calculations in \autoref{ssec:tml_cool_rate}).
        The value of \vent is regulated by the turbulence diffusion and cooling rate within the TML, related to the two regimes discussed above.
        By the conservation of angular momentum, the gas that cooled from the rather static CGM would flow in (\autoref{fig:tml_cartoon}b) at the rate of \dMin, which we estimate in \autoref{ssec:tml_inflow_rate}.
\end{enumerate}

\subsection{The Global Cooling Rate}
\label{ssec:tml_cool_rate}
Under the turbulent-mixing gas cooling scenario, the hot gas would mix and cool in the TML, and then get entrained into the cool phase.
Below, we estimate this global cooling/entrainment rate \dMcool from observations, using the expression of $\dd\dMcool\big/\dd A=\rhocgm\vent$ in \autoref{ssec:tml_expectation}.

We calculate \dMcool within the region that has moment-2 values (\autoref{fig:deconv_result}c), and exclude the inner disk with $r<\Rzs$ and the highly dispersed region with $\sigv>\rnghighlydisperse$, where feedback and fountain possibly play a more important role.
We take the velocity dispersion of diffuse \Hi (\sigv) as \sigturb, and \vtot (\autoref{ssec:vfieldmod}) as the shearing velocity \vs.
At each pixel, an additional factor of $2/\cos{i}$ is multiplied to account for the projection of area and the two sides of \Hi structure.
We estimate the cooling time \tcool as $(5/2)(k\tsb{B}\Tmix)^2/p\Lambda$, where $p$ and $\Lambda$ are the partial pressure of the hydrogen and cooling function at $\Tmix=\qty{2e5}{\K}$, with a metallicity of $0.3Z_\odot$, which is typical for the CGM of an MW-type galaxy such as M51 \citep[e.g.,][]{2015ApJ...800...14M,2017ASSL..430..117L,2017ApJ...837..169P}.

We conduct three sets of estimations of \vent and \dMcool, using both an idealized model and the latest calibrations from hydrodynamic simulations.
Their order-of-magnitude consistency lends support for the estimation, while their differences represent the model uncertainties.
\begin{enumerate}
  \item \emph{Small-perturbation approximation.}
        From the equation of continuity, \citetalias{2019MNRAS.487..737J} obtained the approximate relation $\vent/\vs\sim\hturb/L$.
        Combining \autoeqref{eq:h_sig}, \vent can be estimated as $\vent\sim\sigturb^2\big/\vs$.
        The resultant global cooling rate \dMcool of \qty{2.1}{\Msun\per\yr} could cumulatively build a column density of \qty{8.3e18}{\per\cm\squared} in \qty{\sim300}{\Myr} averaged over the region included in \dMcool calculation.
  \item \emph{The \citetalias{2019MNRAS.487..737J} hydrodynamic calibration.}
        \citetalias{2019MNRAS.487..737J} calibrated a scaling relation of $\vent(\vs, \tcool)$ (their Equation~24) using 3D hydrodynamical simulations.
        \citetalias{2021MNRAS.502.3179T} pointed out that the \citetalias{2019MNRAS.487..737J} calibration was carried out for the weak-cooling regime, and should be scaled down by $\Da^{-1/4}$ when describing the strong-cooling regime.
        We calculate \Da per pixel as $(\hturb/\sigturb\tcool)^{4/3}$ \citepalias{2021MNRAS.502.3179T}, and obtain a global cooling rate of \qty{\sim1.8}{\Msun\per\yr}.
  \item \emph{The \citetalias{2021MNRAS.502.3179T} hydrodynamic calibration.}
        With Equations (53) and~(54) of \citetalias{2021MNRAS.502.3179T}, we derived a global cooling rate of \qty{3.9}{\Msun\per\yr}.
        In this case, the eddy scale \Leddy and the turbulence velocity $u'$ at this scale are required, but are difficult to constrain from observation.
        Therefore, we estimate them using Equation~(55) in \citetalias{2021MNRAS.502.3179T}, with an assumed \Leddy of \qty{100}{\pc} and cool gas sound speed of \qty{15}{\km\per\s}, the typical values that \citetalias{2021MNRAS.502.3179T} used.
\end{enumerate}

\begin{table}
  \footnotesize
  \caption{Global Turbulent-mixing Gas Cooling Rates}
  \label{tab:tml_cooling}
  \iflatexml\else
  \setlength{\tabcolsep}{0pt}
\fi
\begin{tabular*}{\linewidth}{@{\extracolsep{\fill}}cccc}%
  \hline\hline
  Global $\dMcool$           &
  $\vent\sim\sigturb^2/\vs$  &
  \citetalias{2019MNRAS.487..737J}
  Calibration                &
  \citetalias{2021MNRAS.502.3179T}
  Calibration \\
  (\unit{\Msun\per\yr})      &     &     & \\
  \hline
  Modeled \vs                & 2.1 & 1.8 & 3.9 \\
  $\vs=\qty{120}{\km\per\s}$ & 1.8 & 1.9 & 3.8 \\
  $\vs=\qty{150}{\km\per\s}$ & 1.4 & 1.6 & 4.3 \\
  \hline
\end{tabular*}

  \par
  \textbf{Note.}
  The cooling rates are summed over all diffuse-only region, excluding the stellar disk ($r<\Rzs$) and the highly dispersed region.
  Modeled inclination $i$ and shearing velocity \vs are reconstructed using tilted-ring modeling (\autoref{ssec:vfieldmod} and \autoref{app:sec:vfieldmap}).
  Three models are used for the estimation of hot-gas entrainment velocity \vent.
  Additionally, the values assuming a constant $\vs=\qtylist[list-pair-separator=\text{ or }]{120;150}{\km\per\s}$ are reported in the second row.
  For more information, refer to the text in \autoref{ssec:tml_cool_rate}.
\end{table}

We summarize the estimations of global cooling rate in \autoref{tab:tml_cooling}, with additional values using a constant $\vs=\qty{120}{\km\per\s}$ (roughly the median \vtot beyond \Rzs) or \qty{150}{\km\per\s} (similar to the CGM sound speed \cscgm) to circumvent the uncertainty of tilted-ring model.
All of these estimations are comparable to the SFR of \qtysfr in M51 \citep{2019ApJS..244...24L}, indicating that the turbulent-mixing gas cooling may contribute significantly to the fueling of star formation, if they can be efficiently transported inward to the star-forming disk.
We discuss the possible inflow in the next section.

\subsection{The Global Inflow Rate After Entrainment}
\label{ssec:tml_inflow_rate}
The cooling of rotationally lagged CGM onto the \Hi disk reduces the specific angular momentum and induces inflow.
We assume that before condensation, the hot CGM has zero velocity on average, given that it is primarily supported by dispersion and that currently, the CGM velocity field cannot be conclusively modeled from absorption \citep[e.g.,][]{2019ApJ...878...84M,2020ApJ...897..151F,2023MNRAS.523..676W}.
One study reported that the hot gas within \qty{50}{\kpc} around the MW has a rotational velocity as high as \qty{75}{\percent} of that of the stellar disk \citep{2016ApJ...822...21H}.
However, they used \ion{O}{7} line measurements, thus biased toward metal-rich gas and likely related to galactic fountains, rather than to the bulk motion of less-enriched low-spin CGM \citep{1998MNRAS.295..319M}, the latter of which is relevant to the gas accretion we discuss here.

For a circularly rotating ring with the radius \rring, the inflow rate induced by \emph{cooling at this radius} would be
\begin{equation}\label{eq:dMinTML}
  \dMinTML\left(\rring\right) = 2\pi\rring^2\left\langle\frac{\dd\dMcool}{\dd A}\right\rangle
\end{equation}
with an assumed flat rotation curve.
Here, $\left\langle\dd\dMcool\big/\dd A\right\rangle$ is the entrainment rate per area $\rhocgm\vent$ averaged over the ring.

\begin{figure}
  \centering
  \includegraphics{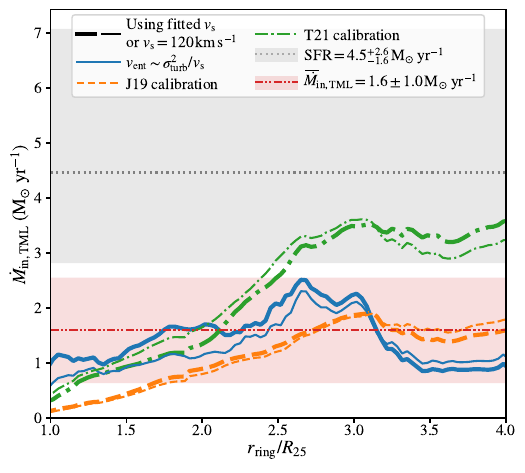}
  \caption{
    The estimated inflow rates due to turbulent-mixing gas cooling at each modeled ring of radius \rring.
    The curves with different line styles represent the results from three different models:
    the continuity model of $\vent\sim\sigturb^2\big/\vs$ (blue solid), \citetalias{2019MNRAS.487..737J} (orange dashed), and \citetalias{2021MNRAS.502.3179T} (green dashed--dotted) calibrations.
    The thin and thick curves correspond to the estimation using the tilted ring-fitted velocity or a constant shearing velocity of $\vs=\qty{120}{\km\per\s}$.
    The value of the SFR is the horizontal gray dashed line, and the averaged value of the curves across the radial range in display is the horizontal red dashed--dotted--dotted line;
    the uncertainties are indicated as the shaded region.
  }
  \label{fig:tml_inflow}
\end{figure}

We calculated this value between \Rzs and $4\Rzs$ using the rings from tilted-ring modeling, shown in \autoref{fig:tml_inflow}.
Averaged over this radial range and over six models, the \bardMinTML is \qty{1.6}{\Msun\per\yr} (red dashed--dotted--dotted line, with a scatter of \qty{1.0}{\Msun\per\yr}), roughly one-third of the SFR\@.
Therefore, the inflow rate, with some scatter, could account for a considerable part of the SFR fueling.
Interestingly, based on interferometric data, the dense-\Hi inflow rate is measured to be $\numrange{\sim0.1}{0.2}\times\text{SFR}$ for the \qty{1e20}{\per\cm\squared} gas at a typical radius of $\numrange{1.5}{3}\Rzs$ \citep{2021ApJ...923..220D}.
Turbulent-mixing gas cooling could serve as an important channel of gas accretion from CGM that replenishes both the outer disk of dense \Hi and the inner star-forming disk.

We note that by plotting \dMinTML as a function of \rring, we do not intend to claim a full description of the continuous radial inflow across the \Hi structures driven by turbulent-mixing gas cooling, but to infer the feasibility of such a picture from an order-of-magnitude and mass-conservation perspective.
If such a gas inflow exists in a less dynamic and quasi-equilibrium system, \dMinTML should remain roughly constant throughout the radial range, and fluctuate around the global cooling rate and SFR.
It is interesting that this prediction is roughly what we find in \autoref{fig:tml_inflow}, which urges the next-step extension to more isolated galaxies.

\subsection{Caveats and Other Possible Effects}
\label{ssec:rp_shock}
During the calculation of cooling and inflow rates, we make many simplifications and approximations.
We adopt the TML scaling relations calibrated from idealized TML simulations by \citetalias{2019MNRAS.487..737J} and \citetalias{2021MNRAS.502.3179T}, who used typical configurations of radiation, temperature, metallicity, etc.~for the CGM of MW-type galaxies.
These two studies did not explicitly consider the role of dust, which, however, is poorly constrained in observations and should not have a significant role beyond the stellar disk, the main region of our analyses.
While the heating and cooling mechanisms are indeed complex, with possibly varying conditions in real galaxies, their overall effects are practically encapsulated into parameters such as cooling coefficients or cooling timescales.
Because the dependence on these cooling parameters has been included in the calibrated scaling relations, which are carried out with CGM-like parameters as mentioned, these scaling relations remain broadly applicable for our case, without being tied to the detailed physical modeling of specific heating and cooling mechanisms.

The entrainment velocity \vent could contribute to the observed values of the moment-2 map, making \sigturb overestimated.
Meanwhile, by using the value of \sigv as \sigturb, \sigturb is underestimated in the weak-cooling regime (\autoref{eq:sig:weak_cool}).
These effects lead to both the over- and underestimation of the cooling rate.
The eddy scale \Leddy is also a source of uncertainty to the estimation based on the \citetalias{2021MNRAS.502.3179T} model.

The tidal forces may cause the \Hi global structure to expand, and slow down the radial inflow rate.
However, for the observed snapshot of M51, the tidal effects on the diffuse \Hi is likely weak, as the current radial extension of \Hi is much larger than the prediction in the simulation of \citet[see \autoref{sssec:comp_simu}]{2010MNRAS.403..625D}.

If the \Hi gas is in the form of small, separate clouds, the ram pressure may slow them down, causing them to flow in \citep[e.g.,][]{2016MNRAS.455.1713H} until they become interlinked.
After that, ram pressure could both strip and compress the diffuse \Hi \citep{1972ApJ...176....1G}, causing it to lag behind dense \Hi as shown in the PV diagram (\autoref{fig:structure_pv}a--c).
Therefore, ram pressure can both assist and harm the growth and inflow of diffuse \Hi.

Finally, taking into account the feedback-driven outflows in the star-forming disk, the total inflow rates onto the disk should be systematically higher than the SFR under a quasi-steady assumption, so as to sustain the star formation.
We thus expect the turbulent-mixing gas accretion to work together with other channels, particularly the galactic fountains within an optical disk, where we exclude for most of the time in this study.
The necessity of having different gas-accretion channels, as well as the separate role of each, has been the topics of recent cosmological simulations \citep{2019MNRAS.490.4786G}.

\subsection{The Turbulent-mixing Cooling in the Context of Gas Accretion and Galaxy Evolution}

MW-type galaxies must accrete gas;
otherwise, they cannot show their current neutral-gas richness after the cosmic-time consumption from star formation \citep{2020ARA&A..58..363P,2020ARA&A..58..157T}.
While merger and stripping of gas-rich satellites cannot make a significant contribution \citep{2013MNRAS.430.1447K}, the cooling and accretion from the virialized CGM (the so-called hot-mode gas accretion) is the most likely origin.

There are several theoretically possible ways for the accretion to take place.
The most extensively studied mechanisms include the precipitation, the cooling flow, and the fountain.
In the precipitation scenario, $\tcool\ll t_\text{dyn}$ (the dynamical time), cool clouds condense out of the hot CGM due to thermal instabilities, and drop onto the galaxy due to the loss of buoyancy \citep[e.g.,][]{2017ApJ...845...80V,2020MNRAS.494L..27G}.
In the cooling flow scenario, $\tcool\sim t_\text{dyn}$, the hot gas flows radially inward due to the loss of pressure support, with radiative cooling balancing the compression heating;
the cooling into the cold phase only happens when it gets close to the disk \citep[e.g.,][]{2022MNRAS.514.5056H}.
In the fountain scenario, metal-enriched cool gas is launched by supernovae perpendicular to the disk and then rains back, while CGM cooling is boosted through the mixing with it \citep[e.g.,][]{2017ASSL..430..323F}.

In the current study, we advocate an alternative scenario of disk--CGM interface cooling, where the tidal structures are counted as the extension of the \Hi disk, and the turbulent mixing serves as the primary driver.
The key requirement for this scenario is an extended \Hi structure with subsonic shearing motion relative to the CGM\@.
Since most star-forming galaxies have a large \Hi disk rotating at a velocity close to the virial one, this scenario can be applied to general star-forming galaxies.

The disk--CGM interface cooling has clear links to the other three scenarios.
The precipitation and cooling flow are like the preparing process, bringing gas close to the disk--CGM interface for further turbulent-mixing gas cooling.
The cooling at the interface between the launched clouds and hot gas is an essential part of the galactic fountain model, whose localized thermodynamic physics is very close to the disk--CGM interface cooling.
The former increases interface with the CGM initially through ejecting outflows, while the latter through tidal interaction or simple mechanisms sets the \Hi size--mass relation.

Meanwhile, its difference from the other three scenarios is also clear.
The proposed cooling process happens at the interface of an existing \Hi disk or structure with the CGM, without the necessity of requiring the final success of the other three CGM cooling scenarios \citep[e.g., not all precipitated clouds survive the journey from large distances onto the disk,][]{2020MNRAS.492.1841L}.
It is a promising way to maintain the gas-richness of a galaxy that is already gas rich, and sustain the existing active star formation.

At cosmic noon, turbulent mixing potentially enhances the cold-mode accretion efficiency \citep{2024MNRAS.532.2965A}.
In this scenario, the shearing happens between the cold stream and hot CGM, and could be the high-redshift counterpart of our case.

\section{Summary}
\label{sec:summary}
Using a new MEM pipeline, we succeed in jointly deconvolving the single-dish and interferometric \Hi cubes from FAST FEASTS and VLA THINGS of the extragalactic M51 system.
At a spatial resolution of \qtyVLAfwhm (\qtyphysVLAfwhm at \qty{8.0}{\Mpc}), the \los column density sensitivity is \qtyNHilim[\sim] at $2\sigma$ confidence.
With the deconvolved cube and moment maps, we obtain the following results:
\begin{enumerate}
  \item The high-\NHi structures are consistent with the prediction of a previous hydrodynamical simulation \citep{2010MNRAS.403..625D}, but the newly detected low-\NHi structures are much more extended.
        It is thus necessary to take the CGM--ISM interaction into consideration.
  \item Roughly half (\qty{48}{\percent}) of the \Hi mass in M51 is diffuse, and \qty{59}{\percent} of this diffuse \Hi is not associated with any dense \Hi (or is ``diffuse-only''), covering \qty{89}{\percent} of the \los sample.
  \item For diffuse-only \Hi, the values of \sigv and \NHi show a power-law relation of $\sigv\propto\NHi^{0.5}$.
  \item The \Hi gas outside of the galaxy optical radius (\Rzs) is found to probably survive photoionization from the UV background, but may be significantly influenced by thermal evaporation under the \citet{1977ApJ...211..135C} model.
\end{enumerate}

With these observational results, we propose that in this system, turbulent-mixing gas cooling could be an important and previously overlooked hot gas cooling and accretion channel.
In this scenario, the vast CGM--ISM interface is generated by tidal interaction, where the shearing between them produces TMLs and induces cooling.
We make the following developments in linking the TML theories and simulations with \Hi observations.
\begin{enumerate}
  \item We design ensemble measurements and estimators for the attributes of TML-related \Hi gas, including the thickness, velocity dispersion, and pressure.
        The properties of M51 diffuse \Hi, particularly the \sigv--\NHi relation, are consistent with the scenario that they are primarily the \Hi that has condensed from the TML\@.
  \item We estimate the rates of turbulent-mixing gas cooling and the subsequent gas inflow using three different recipes.
        These rates are around one-third of the M51 SFR\@.
        Such a cooling could accumulate an averaged \Hi column density of \qty{\sim1e19}{\per\cm\squared} in \qty{\sim300}{\Myr}, potentially accounting for a considerable part of cold gas replenishment.
\end{enumerate}

This work illustrates the possibility of probing the multiscale structures, kinematics, and processes at ISM--CGM interface through \Hi emission observations.
The proposed turbulent-mixing gas accretion is worth investigation for more-general \Hi-rich galaxies.
For a FEASTS sample of 35 moderately inclined \Hi-rich galaxies, the \Hi disks down to the $\NHi\sim\qty{1e18}{\per\cm\squared}$ level on average extend to \qty{70}{\kpc} or \num{0.3} times virial radius \citep{2025ApJ...980...25W}, providing a large area of shearing interface with the hot CGM\@.
Previously, these $\NHi\sim\qty{1e18}{\per\cm\squared}$ regions could only be detected from low-ion metal absorption along selected \los{}s.

Our observation and analyses on M51 favor the picture that the \Hi gas around this \Hi-rich merging system distributes as a relatively contiguous and large structure extending into the CGM\@.
At the ISM--CGM interface, the proposed turbulent-mixing gas accretion offers a possible explanation for the cold-gas enhancement observed statistically in post-merger galaxies \citep[e.g.,][]{2015MNRAS.448..221E,2018MNRAS.478.3447E,2019ApJ...870..104S}.

We expect systems similar to M51 can be more common in the early Universe where mergers are more frequent and gas-rich.
If so, a large fraction of \qty{\sim1e18}{\per\cm\squared} gas detected by absorptions could be related to tidally deformed \Hi structures, particularly at the inner part of the halo, rather than being the projection of a 3D sea of distinct clouds, or forming a number of major filaments.
Also, the possible turbulent-mixing gas cooling may contribute more significantly to the mass assembly of these high-redshift galaxies.
More future combinations of FEASTS and archival or new interferometric data could help to further constrain this key component of gas accretion.

\section*{Acknowledgments}

We thank the anonymous referee for providing constructive and helpful comments.
We thank Bi-Qing For, Gregory J. Herczeg, Jin Koda, Thijs van der Hulst, Junfeng Wang, Simon Weng, and Tobias Westmeier for useful suggestions and discussions.
We asked Google Gemini (1.5 Flash, accessed on 2024 October 29) for suggestions on the grammar and expression of the Introduction.

J.W. acknowledges research grants and support from the Ministry of Science and Technology of the People's Republic of China (No.\ 2022YFA1602902), the National Natural Science Foundation of China (No.\ 12073002), and the China Manned Space Program (No.\ CMS-CSST-2021-B02).
S.J. acknowledges support from the National Natural Science Foundation of China (Nos 12133008, 12192220, and 12192223), the China Manned Space Program through its Space Application System, and the National Key Research and Development Program of China (No.\ 2023YFB3002502).
N.M. acknowledges support from Israel Science Foundation (ISF) grant 3061/21 and US--Israel Binational Sciences Foundation (BSF) grant BSF~2022281.
S.H.O. acknowledges support from the National Research Foundation of Korea (NRF) grant funded by the Korea government (Ministry of Science and ICT: MSIT) (No.\ RS-2022-00197685).

This work has used the data from FAST (\url{https://cstr.cn/resolver?identifier=31116.02.FAST}).
FAST is a Chinese national mega-science facility, operated by the National Astronomical Observatories, Chinese Academy of Sciences (NAOC)\@.

\facilities{FAST:500m, NED, VLA.}

\software{Astropy 5.2.2 \citep{2022ApJ...935..167A}, BBarolo 1.7 \citep{2015MNRAS.451.3021D}, CASA 6.5.3 \citep{2022PASP..134k4501C}, Karma 1.7.25 \citep{1996ASPC..101...80G}, mosmem 1.10 and other MIRIAD utilities \citep{1995ASPC...77..433S}, NumPy 1.24.3 \citep{2020Natur.585..357H}, Python 3.10.12, pingouin 0.5.4 \citep{2018JOSS....3.1026V}, pvextractor 0.3 \citep{2016ascl.soft08010G}, regions 0.8 \citep{2023zndo..10144914B}, SciPy 1.10.1 \citep{2007WRR....43.3423V,2020NatMe..17..261V}, SoFiA 2.5.0 \citep{2021MNRAS.506.3962W}.}

\appendix

\section{Improving the Continuum Subtraction of the VLA Data in the Visibility Domain}
\label{app:sec:contsub}
The limited velocity range of THINGS data (from \qtyrange[range-phrase={ to }]{214.9}{704.5}{\km\per\s}) leads to an improper continuum subtraction at $v\ge\qty{658.1}{\km\per\s}$ \citetext{cf.~the VLA observation of \citealp{1990AJ....100..387R} with a broader spectral window}.
It mainly influences a small region of \qty{\sim13}{arcmin^2} to the north of M51b (see the contour on the channel maps in \autoref{app:sec:chan_map}).

We thus added back the missing interferometric flux in that region using the data from a shallower VLA observation by \citet{1990AJ....100..387R}.
With a beam of $\ang{;;34.7}\times\ang{;;27.5}$, the \citet{1990AJ....100..387R} data have an rms level of \qty{0.69}{\mJy\per\beam}.
We clean the data of \citet{1990AJ....100..387R} and THINGS%
\footnote{
  When doing the correction, we taper the THINGS cube so that it has the same clean beam as the \citet{1990AJ....100..387R} data.
  The rms level becomes \qty{0.57}{\mJy\per\beam} after this tapering.
}
with MIRIAD \task{clean} down to their respective $1\sigma$ levels.
We obtain the difference of their cleaned models within this small region, and convert the difference into visibility, which is added onto the original THINGS visibility.

We note that this relatively narrow velocity range does not introduce a significant loss of \Hi flux.
Only \qty{0.16}{\percent} of the M51 flux detected by FEASTS is located beyond the velocity of \qty{704.5}{\km\per\s}, the maximum velocity observed by THINGS\@.

\section{The Joint Deconvolution Pipeline}
\label{app:sec:pipeline}
In this appendix, we provide the details of the joint deconvolution pipeline introduced in \autoref{ssec:pipeline}.

\paragraph{Data Preparation}
The interferometric data are the dirty cube \intim described in \autoref{ssec:vla_data}.
The dirty beam image \intb is twice as large as the cube, of $\num{3600}\times\num{3600}$ pixels.

For the single-dish cube \sdim, we use the FAST data after flux cross-calibration (scaled down by a factor of \num{1.15}) with the interferometric data as described in \citetalias{2024ApJ...968...48W}, and regrid them into the coordinate system of \intim.
The World Coordinate System calibration is similar to \citetalias{2024ApJ...968...48W}, but is done within the central $\ang{;;600}\times\ang{;;600}$ box instead, which fully covers the brightest inner \Hi disk.
The world-coordinate reference pixel values, \para{CRPIX1}/\para{2}, of FAST cube are increased by \num{-0.15} and \num{-0.2}.
The single-dish beam \sdb is the average of the 19-beam FAST receiver beams described in Appendix~B.2 of \citetalias{2024ApJ...968...48W}, after necessary coordinate projections.

The mask for the MEM deconvolution is taken as the mask of the FAST cube generated using SoFiA as described in Section~2.2 of \citetalias{2024ApJ...968...48W}.
The regridded single-dish cube \sdim has an rms of $\sigsd=\qty{0.489}{\mJy\per\bfast}$ beyond the mask, lower than the one (\qty{0.80}{\mJy\per\bfast}) reported in \citetalias{2024ApJ...968...48W} due to the binning during regridding.
Only within this deconvolution mask will the MEM deconvolution subroutine generate the nonzero pixels of the solution \modim, calculate the entropy $H$, and measure the residual rms.

The cubes of \intim and \sdim and the deconvolution mask are then separated into individual channels.
The following deconvolution procedure is then run per channel as described below.
For simplicity, these single-channel images are also referred to as \intim and \sdim.

\paragraph{The Procedures}
The major steps of the single-channel joint deconvolution iterations are as follows.
In each of these iterations (labeled as $n$th), the MEM routine \task{mosmem} is invoked once with the same \intim and \sdim, and different reference images \defim[n].
\task{Mosmem} derives the best model \modim[n] for the given \defim[n] and threshold \thres[n]{}, as described in \autoref{ssec:pipeline}.

We always set the rms convergence threshold for single-dish map as $\thres[n]{sd}=\thres[n]{int}\unit{\bfast}\big/\unit{\bvla}$,%
\footnote{
  Thus, both \thres[n]{} have the same value when they are converted to \unit{\Jy\per\pixel}.
  In other words, the interferometric and single-dish images share the same physical rms criterion.
}
and both \thres[n]{}'s have a tolerance of \qty{5}{\percent}.
In each iteration, the initial guess for \modim[n] is always a flat estimate filling the deconvolution mask with the total flux of \sdim.
\begin{enumerate}
  \item \emph{First iteration.}
        In the first call of MEM deconvolution, \defim[1] is set to be \sdim, and the interferometric rms threshold $\thres[1]{int}=1.5\sigint$.
  \item \emph{Generate the new reference image.}
        \label{mosmem:adjust}
        The output model \modim[n] (after primary-beam (PB) correction) and residuals of the $n$th iteration are combined to produce the new reference \defim[n+1] for the next iteration.
        We search for residual peaks that are potential signals, and add them onto \modim[n] as \defim[n+1], thus increasing the deconvolution quality and speeding up the convergence, which shares the same concept as the peak-searching step between iterations in traditional CLEAN \citep{1974A&AS...15..417H,1980A&A....89..377C,1984A&A...137..159S}.

        For the interferometric residual $\intres[n]=\intim-\intb*\left(\text{PB}\times\modim[n]\right)$, we select out the pixels where both \intres[n] and $\text{PB}\times\modim[n]$ are high enough relative to the interferometric noise level \sigint.%
        \footnote{
          Here, the \modim[n] is compared with \sigint instead of $\sqrt{\sigint\sigsd}$, because the goal is to select out the high S/N region relative to the interferometric residual that will be added back.
        }
        We require the product of these two S/N values to be higher than $3^2$, so as to include pixels where at least one S/N is very significant.
        A hyperbolic tangent ($\tanh$) function is used to generate a smooth weight map and thus to reduce the clumpiness of new reference map.

        In a similar way, the pixels of single-dish residual $\sdres[n]=\sdim-\sdb*\modim[n]$ are selected out with a threshold of $3^2$.
        We use two sets of criteria, (a)~comparing \sdres[n] and \modim[n] with \sigsd and $\sqrt{\sigint\sigsd}$ (converted to the same beam area) and (b)~comparing \sdres[n] and the smoothed \intres[n] with \sigsd and $\sigint\big/\sqrt{1+3^2}$, where the Gaussian smooth kernel has three times longer major and minor axes than the clean beam.
        The motivation of criterion~b is to find the region where both data sets have significant positive residuals.
        We compare the weights from these two criteria, and only use the larger one.

        When added onto \modim[n], the weighted \intres[n] is PB-corrected, and is scaled down to account for the difference in the dirty and clean beam areas \citep{1995AJ....110.2037J}.
        To avoid repetitively adding fluxes, we subtract the weighted \intres[n] from the weighted \sdres[n] values after convolving it with the single-dish beam \sdb, and replace the negative pixels in the resultant \sdres[n] with \num{0}.
  \item \emph{The $(n+1)$th iteration.}
        \label{mosmem:rerun}
        We calculate the interferometric residual rms \sigintout[n] \emph{outside} the joint deconvolution mask of the previous residual \intres[n].
        If $\sigintout[n]<\thres[n]{int}$, we update the \thres[n+1]{int} to be \sigintout[n];
        otherwise, \thres[n+1]{int} is kept as \thres[n]{int}.
        Then, the MEM routine is called with the new interferometric rms threshold \thres[n+1]{int} and the new reference image \defim[n+1] from step~\ref{mosmem:adjust}.
  \item \emph{Stop criterion.}
        \label{mosmem:stop}
        The iteration is repeated for at least eight times.
        A channel is considered fully deconvolved, and thus the iteration stops, if the interferometric residual rms \emph{within} the deconvolution mask is lower than the one out of the mask, \sigintout[n].
        We choose this criterion because the major bottleneck of joint deconvolution is the interferometric noise.
        Otherwise, steps \ref{mosmem:adjust} and~\ref{mosmem:rerun} would be repeated.
  \item \emph{The solution for this channel.}
        The stop criterion of step~\ref{mosmem:stop} only considers the interferometric residual.
        To quantify the overall quality of the output models, we use the average relative residual level \goodness, calculated as $\sqrt{\goodness[int]\goodness[sd]}$, where \goodness[int] (or \goodness[sd]) is the interferometric (single-dish) residual rms \emph{within} the deconvolution mask and has been normalized by \sigint (\sigsd).
        For each channel, the output of the iteration with the lowest \goodness is automatically selected as the final solution.
        We note that two out of 88 channels have a very small deconvolution mask, and thus do not meet the stop criteria within 20 iterations.
        For them, the final iteration model is used.
        We also manually select the best iteration for two other channels for a better overall residual map.
\end{enumerate}

\paragraph{Final Data Products}
When all channels have finished the above iterations, we convolve each of the output models (\modim) with the restoring beam that is chosen as interferometric clean beam (\cleanb), and combine all of them into one single cube.
The medians and \qty{90}{\percent} distribution intervals of the relative residual levels \goodness, \goodness[int], and \goodness[sd] are \iflatexml$1.54^{+0.97}_{-0.43}$, $1.11^{+0.04}_{-0.12}$, and $2.1^{+3.4}_{-1.0}$\else\numlist[parse-numbers=false]{1.54^{+0.97}_{-0.43};1.11^{+0.04}_{-0.12};2.1^{+3.4}_{-1.0}}\fi, respectively.

\section{A Justification of the Joint Deconvolution Procedure}
\label{app:sec:pipeline_justification}
One typical concern about the deconvolution is overdeconvolution.
A reasonable expectation of deconvolution is that, similar to the traditional clean used in deconvolving dirty images, it can only sharpen the features well-resolved at the existing resolution.
Indeed, one important role of the information entropy $H$ is to suppress overdeconvolution, particularly when the FAST image, which has a large but relatively simple beam, is used as the initial reference image \defim[1] (i.e., the zero-point of $H$) to calculate the $H$.
Hence, the model \modim[1] by construction should be relatively smooth, at the expense that the localized peaks may be overlooked (i.e., insufficient deconvolution).
At early iterations, such a disadvantage of underdeconvolution is manifested as the clear and sharp patterns
remaining in the residual images \resim[n]{int,sd}.

To compensate for the insensitivity at small scales, one key feature of our pipeline is to iteratively decide whether the connected positive pixels in \intres[n] contain statistically significant signals, and thus whether they should be added back to make a new reference image \defim[n+1].
This step is very similar to the step between multiscale clean iterations \citep{2008ISTSP...2..793C}.
The difference (advantage) here is that the FAST signals locally help to verify the THINGS residual patterns with only a moderately S/N, which would not be considered statistically significant just by themselves.
By construction, the THINGS image contributes the most information where it has sufficient S/N, the FAST image takes over where the THINGS image only contains noise, and these two images determine the model together in the intermediate region.
Putting together, the whole procedure is designed to achieve a model that is incompletely but sufficiently deconvolved at the given resolution of raw data.

In this appendix, we discuss the accuracy of our joint deconvolution pipeline from several different points of view.
\autoref{app:ssec:radial_qualitative} gives a qualitative discussion on the PB attenuation.
A mock test quantifies the accuracy of \NHi and \sigv at the effective resolution of restoring beam (\qtyVLAfwhm) in \autoref{app:ssec:recovery}.
We estimate the instrumental resolution directly from the raw-data noise in \autoref{app:ssec:noise_spec}.
Finally, we compare the jointly deconvolved moment-0 map to the data before joint deconvolution and to the result of linear combination in \autoref{app:ssec:mom0_comp}.

\subsection{Influence of the Radially Varying Uncertainties}
\label{app:ssec:radial_qualitative}
One complicating factor is the PB-attenuation effect on the THINGS image, which increases the noise at large radius and thus obscures the THINGS signal.
However, by construction, the THINGS noise only propagates to the regions where THINGS signal is significant, i.e., where PB attenuation is not strong.
There, the THINGS data actually \emph{lower} the systematic uncertainty from the smearing of the FAST beam.
Furthermore, for M51, all of the pixels with $\NHi>\qtyNHilim[]$ (the $2\sigma$ detection limit from \autoref{app:ssec:recovery}) have a PB-attenuation factor \num{>0.5}, indicating that the role of PB is very limited.

However, a following major worry is that the output image (model convolved with the THINGS clean beam, $\cleanb *\modim$) has increasingly lower resolution at larger radius.
Within the scope of this study, the joint-deconvolution result is acceptable, as long as the statistical trends and relations of \los \NHi and \sigv are relatively robust against these uncertainties.
Here, we support this argument with metal experiment, and mock tests on the accuracy of the recovered \NHi and \sigv is given in \autoref{app:ssec:recovery}.

For the value of \NHi, \citetalias{2024ApJ...973...15W} demonstrated that the simple technique of linear combination \citep{2002ASPC..278..375S} already works surprisingly well in statistically recovering its true value at a large radius, or the low \NHi values far below the interferometry detection limit.
The stochastic and systematic uncertainties are \qty{<0.1}{\dex}.
The key is that by incorporating the interferometric data, we remove the additional light in the FAST-only region that is scattered from the high-\NHi regions.
The joint-deconvolution procedure effectively improves the removal of scattered light, and should achieve an even more-reliable measurement of \NHi.

As for the kinematical measurements, at a large radius of $r$, the velocity smearing is insignificant when the velocity field $v$ is relatively smooth.
To the first order, the smearing effects can be estimated as $\delta\sigv= v\sin i\sin\phi\,\delta\phi$, where $\phi$ is the azimuthal angle starting from the major axis of a tilted ring, and $\delta\phi=0.5\text{FWHM}\big/r\sqrt{\sin^2\phi+\cos^2i\cos^2\phi}$ is the azimuthal angle spanned relative to the center by a resolution element of half the FAST beam.
Assuming that $v=\qty{100}{\km\per\s}$ (roughly the median velocity of our fitted \vtot profile), $i=\ang{55;;}$ at $r=\ang{;15;}$ (\qty{35}{\kpc} or $3\Rzs$ for M51) and the azimuthally averaged velocity smearing is $\langle\delta\sigv\rangle=0.5\text{FWHM}/r\times 2vi/\pi=\qty{6.6}{\km\per\s}$, and the maximum value is $\delta\sigma_{v,\text{max}}=v\sin i\times0.5\text{FWHM}/r=\qty{8.8}{\km\per\s}$ at $\phi=\ang{90;;}$, both of which are lower than most of the measured \sigv at that radius (see \autoref{fig:n_h}).
These deductions support our statistical analyses of \NHi and \sigv based on the joint-devolution output.

\subsection{The Mock Test on the Recovery of Column Densities and Dispersions}
\label{app:ssec:recovery}
We largely follow the procedure in \citetalias{2024ApJ...968...48W} and \citetalias{2024ApJ...973...15W} to generate the mock cube of an \Hi disk with the size of $\RHi=\ang{;6.5;}$ and the inclination of \ang{45;;}.
The angular size of this mock \Hi disk is larger than M51 \citepalias[$\RHi=\ang{;6.0;}$,][]{2024ApJ...968...48W}, and thus the PB attenuation more significantly impairs the recovery of \Hi structures in our mock tests than in the real case.

The mock disk has a radial distribution and power spectral slope typical for real galaxies.
However, those two works did not consider the velocity axis, and here, we improve them by generating the smooth part of the disk with BBarolo.
The rotation curve is set to gradually increase from \qtyrange[range-phrase={ to }]{0}{200}{\km\per\s} within $r=\ang{;;120}$ and then flattens, and the velocity dispersion drops from \qtyrange[range-phrase={ to }]{35}{5}{\km\per\s} across the radial range.
No warping, \PA changing, nor radial velocity is included.
Following \citetalias{2024ApJ...968...48W}, we conduct the mock observations with the settings close to those of FEASTS and THINGS\@.
The mock VLA visibility is generated using \para{casa.simobserve}.
We jointly deconvolve the mock FAST cube and the VLA mock dirty cube as for M51 in this paper.

\begin{figure}
  \centering
  \includegraphics{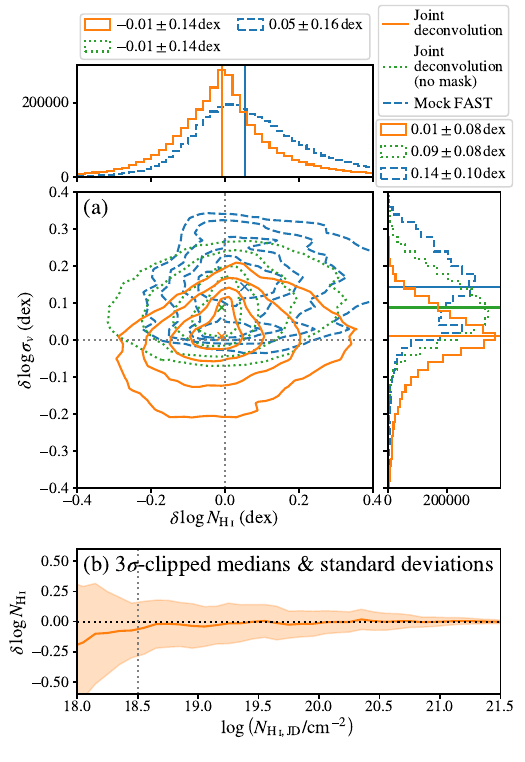}
  \caption{
    The accuracy of the column density (\NHi) and \los velocity dispersion (\sigv) reconstructed by our joint deconvolution pipeline, tested on the mock \Hi cube (see text of \autoref{app:ssec:recovery}).
    (a)~The logarithmic deviations from the mock input of the mock FAST cube (blue dashed lines) and jointly deconvolved results.
    The \sigv values of orange solid and green dotted contours are measured with and without SoFiA mask, respectively.
    The 2D-distribution contours enclose \qtylist{90;75;50;25}{\percent} of the pixels.
    The $3\sigma$-clipped medians of the 1D distributions are indicated as solid lines (and the crosses on the 2D map).
    The corresponding $3\sigma$-clipped standard deviation is also reported in the legend.
    (b)~The logarithmic deviation of jointly deconvolved \NHi plotted against the jointly deconvolved value (\NHi[JD]).
    The solid line is the $3\sigma$-clipped median value, and shaded area indicates the clipped standard deviation.
    The dotted vertical line indicates the \los \NHi[JD] threshold of $2\sigma$ confidence, \qtyNHilim[\sim].
  }
  \label{fig:mock_accu}
\end{figure}

We compare the measurement of the mock FAST cube and jointly deconvolved results with that of the input model in \autoref{fig:mock_accu}(a).
Statistically, both the \NHi and \sigv are recovered by our joint deconvolution pipeline remarkably well, with the $3\sigma$-clipped median deviations of \qty{0.01}{\dex}.
There is no clear correlation between the deviations of two parameters.
The clipped scatters are both of \qty{\sim0.1}{\dex}.
In contrast, the values of \NHi and \sigv measured from the mock FAST cube are both overestimated with larger deviations and scatters.

Our mock test shows that when we calculate the moment-2 map of the joint deconvolution result, a cube mask is necessary, otherwise the \sigv would be systematically overestimated by \qty{0.09}{\dex} due to the noise in the spectral wings.
The cube mask for the moment-2 map is generated using SoFiA~2.
We add onto the cube Gaussian noise of $\sqrt{\sigint\sigsd}$ at the resolution of the restoring beam, before inputting the noised cube to SoFiA\@.
We use a SoFiA smoothing kernel that is \ang{;;30} wide spatially and three channels wide (\qty{15.5}{\km\per\s}) spectrally.
The detection and reliability thresholds are $2\sigma$ and \num{0.99}, respectively.
Only the \los{}s with moment-2 values after we apply the SoFiA mask are counted in \autoref{fig:mock_accu}(a).

For the statistical accuracy of \NHi, we find that it is better not to apply the cube mask when generating the moment-1 map of the joint deconvolution result.
The cube mask would remove the fluxes in the spectral wing, introducing an underestimation of \NHi of \qty{0.02}{\dex} (not shown in \autoref{fig:mock_accu}a).
\autoref{fig:mock_accu}(b) shows that when the jointly deconvolved \NHi[JD] is higher than \qtyNHilim[\sim], the logarithmic deviation has a scatter smaller than $1/(2\ln 10)=\qty{0.22}{\dex}$, i.e., a confidence level of $2\sigma$ for \los \NHi measurements.

Therefore, in this study, the M51 moment-2 map of the jointly deconvolved cube is generated with a SoFiA cube mask, while no mask is applied when we generate the moment-0 and moment-1 maps.

\subsection{The Resolution at Different Column-density Detection Levels}
\label{app:ssec:noise_spec}
In this appendix, we estimate the best instrumental resolution that our joint deconvolution pipeline could recover at a given \NHi detection level.
While for a single-dish-only or interferometry-only data set, its beam shape and residual rms level determine the resolution and detection level, our joint deconvolution pipeline takes two different beams and produces two different residual maps.
Thus, its relation between resolution and final noise level is not self-evident.

To simplify the question, we first omit the details of the pipeline, thereby deciding the limit set by the data themselves.
An ideal combination of interferometric and single-dish data is expected to maximize the S/N at each spatial frequency $k$ in the Fourier $u$--$v$ space.
We model such a combination as weighted average in the $u$--$v$ space, using the same weighting scheme as in \citet[Section~4.1]{2019PASP..131e4505K}.
The averaged dirty beam and noise spectra are
\begin{align}
  \tilb[avg]                & = \alpha\times\left(
  \left|\tilb[int]\big/\tilsig[int]\right|^2
  + \left|\tilb[sd]\big/\tilsig[sd]\right|^2
  \right),\label{eq:bavg}                          \\
  \left|\tilsig[avg]\right| & = \alpha\times
  \sqrt{
    \left|\tilb[int]\big/\tilsig[int]\right|^2
    + \left|\tilb[sd]\big/\tilsig[sd]\right|^2
  },\label{eq:sigavg}
\end{align}
where the constant factor $\alpha$ is chosen to normalize the peak of $b\tsb{avg}$, the averaged dirty beam, as \num{1}.
Here, we use the symbols \tilb and \tilsig to denote the $u$--$v$ spectrum of beam and noise, respectively, and add subscripts to differentiate the interferometric (int), single-dish (sd), and averaged (avg) ones.

\begin{figure}
  \centering
  \includegraphics{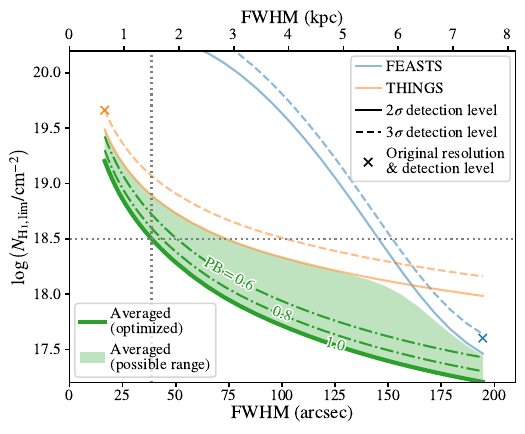}
  \caption{
    The \NHi detection levels allowed by different resolutions for the combination of FEASTS and THINGS observations, modeled as weighted average in the Fourier $u$--$v$ space.
    The thick green line indicates the $2\sigma$ detection limit if the S/N at each spatial frequency is optimized, while the green shaded region gives the whole possible range.
    The two green dashed--dotted lines correspond to the cases where THINGS data are attenuated by PB\@.
    For the $2\sigma$ \NHi limit determined from mock test (\qtyNHilim, horizontal gray dotted line, \autoref{app:ssec:recovery}), the vertical gray dotted line gives the best combined instrumental resolution, \qtyFWHMlim (\qtyphysFWHMlim).
    For THINGS and FEASTS, the solid and dashed lines represent $2\sigma$ and $3\sigma$ limits, respectively.
    For reference, the resolution and detection limits of original FEASTS (blue) and THINGS (orange) data are the two crosses, and the solid (dashed) lines in corresponding color are their relations between the detection level and resolution.
  }
  \label{fig:detectlim}
\end{figure}

\begin{figure}
  \centering
  \includegraphics{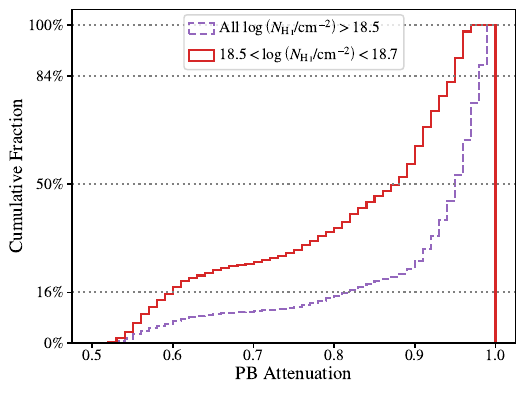}
  \caption{
    The cumulative distributions of PB-attenuation factors.
    The red solid line is the distribution of pixels within \qty{0.2}{\dex} above the $2\sigma$ detection level of \qtyNHilim, and the purple dashed line is that of all pixels above the detection level.
  }
  \label{fig:pb_dist}
\end{figure}

\begin{figure*}
  \centering
  \includegraphics{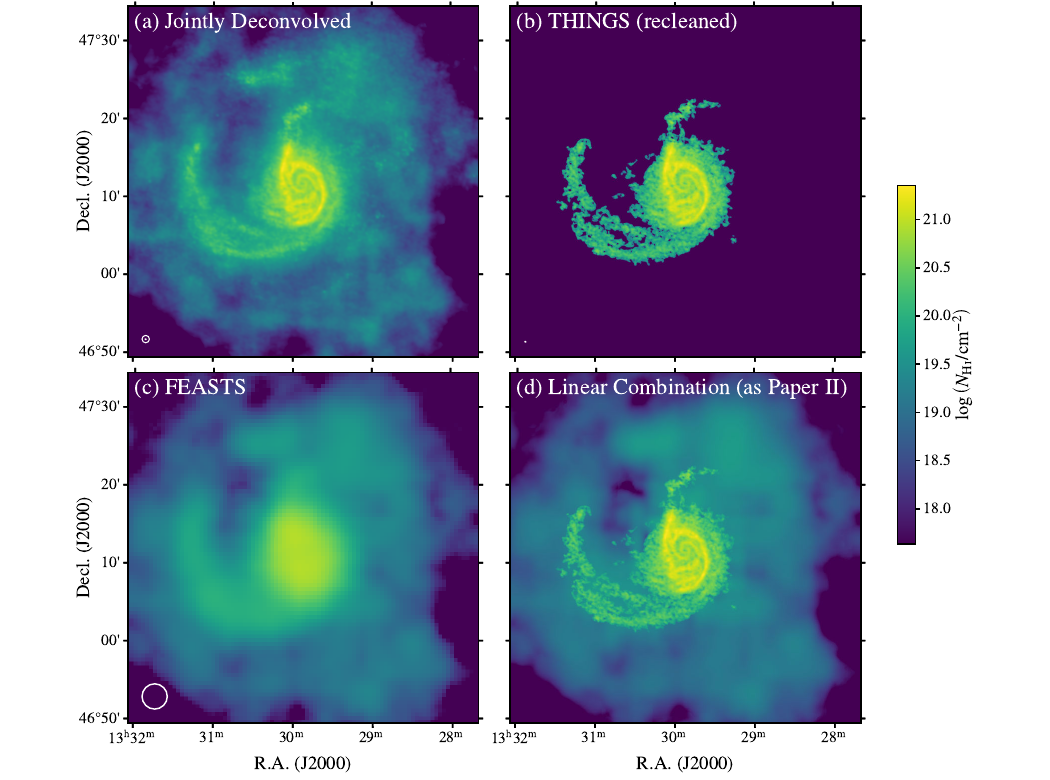}
  \caption{
    A comparison of the moment-0 maps of different data and/or methods under the same color coding.
    (a)~From the new joint deconvolution method.
    (b)~The recleaned THINGS moment-0 map.
    (c)~FEASTS single-dish moment-0 map.
    (d)~The moment-0 map linearly combining FEASTS and THINGS data, using the procedure of \citetalias{2024ApJ...973...15W}.
    In the lower left corner of panel~(a), the open white circle displays the diameter of the instrumental resolution (\qtyFWHMavg) at $\NHi\sim\qtyNHilim[]$, and the white ellipse within is the restoring beam (\VLAcleanbeam).
    The actual resolution transits between these two limits, reaching the best one when $\NHi>\qty{1.6e19}{\per\cm\squared}$.
    The corresponding beam of panels (b) and~(c) is given at the lower left corner as a white ellipse.
  }
  \label{fig:mom0_comp}
\end{figure*}

We transform \tilb[avg] and \tilsig[avg] back to the image domain, and taper them with Gaussian kernels of varying sizes.
From FWHMs of tapered $b\tsb{avg}$'s and rms levels of tapered $\sigma\tsb{avg}$'s, we obtain the lowest $2\sigma$ \NHi detection levels \NHi[lim,avg] allowed by each resolution, shown in \autoref{fig:detectlim} as the green thick line.
The $2\sigma$ detection limits of FEASTS (THINGS) at tapered resolutions are the blue (orange) solid lines, which we denote as \NHi[lim,sd] and \NHi[lim,int], and the $3\sigma$ limits are the dashed lines.
The original resolution and limits are plotted as two crosses.

In the above analyses, we have used a \emph{local-optimization} model assuming that the S/N at each spatial frequency $k$ is maximized, representing the best case of data combination.
On the other hand, if only a \emph{global optimization} is applied across all $k$, the expected detection level would be $1\Big/\sqrt{\NHi[lim,int]^{-2}+\NHi[lim,sd]^{-2}}$, representing the loosest case.
These two cases provide a rather narrow constraint for the possible output \NHi[lim] varying as a function of resolution, plotted as the green shaded region in \autoref{fig:detectlim}.
For the $2\sigma$ \NHi limit of \qtyNHilim (gray dotted line) obtained in \autoref{app:ssec:recovery}, the instrumental resolution range is \qtyFWHMlim--\qtyFWHMlimup[] (\qtyphysFWHMlimrange).
We take their average value (\qtyFWHMavg) as the final estimation for the instrumental resolution.

Although our pipeline conducts joint deconvolution in the image domain, the actual resolution should be closer to the best case (\qtyFWHMlim) than to the loosest case due to the following reasons:
(1)~Between iterations, we correct the reference image with not only the single-dish residual map \resim{sd} and interferometric residual map \resim{int}, but also a smoothed \resim{int}.
This procedure uses information at three spatial frequencies, effectively improving the sampling of the $u$--$v$ space.
(2)~We visually inspect the interferometric residual maps to check the deconvolution convergence, thereby confirming that no obvious moderate- to large-scale features are left in the residual.

We have so far ignored the PB attenuation, which decreases the S/N of interferometric data with a large distance to the image center, but is actually not significant for M51.
In \autoref{fig:pb_dist}, we plot the cumulative distributions of \los with $\log\NHi$ between \numlist{18.5;18.7} (or above \num{18.5}) as a function of the PB attenuation factor, and more than \qty{84}{\percent} of the pixels have a PB-attenuation factor above \num{0.6} (or \num{0.8}).
The corresponding relations between the resolution and detection level with the interferometric noise scaled up by $0.6^{-1}$ (or $0.8^{-1}$) are the green dashed--dotted lines in \autoref{fig:detectlim}, which do not deviate significantly from the line without PB attenuation.

\begin{figure*}
  \centering
  \includegraphics{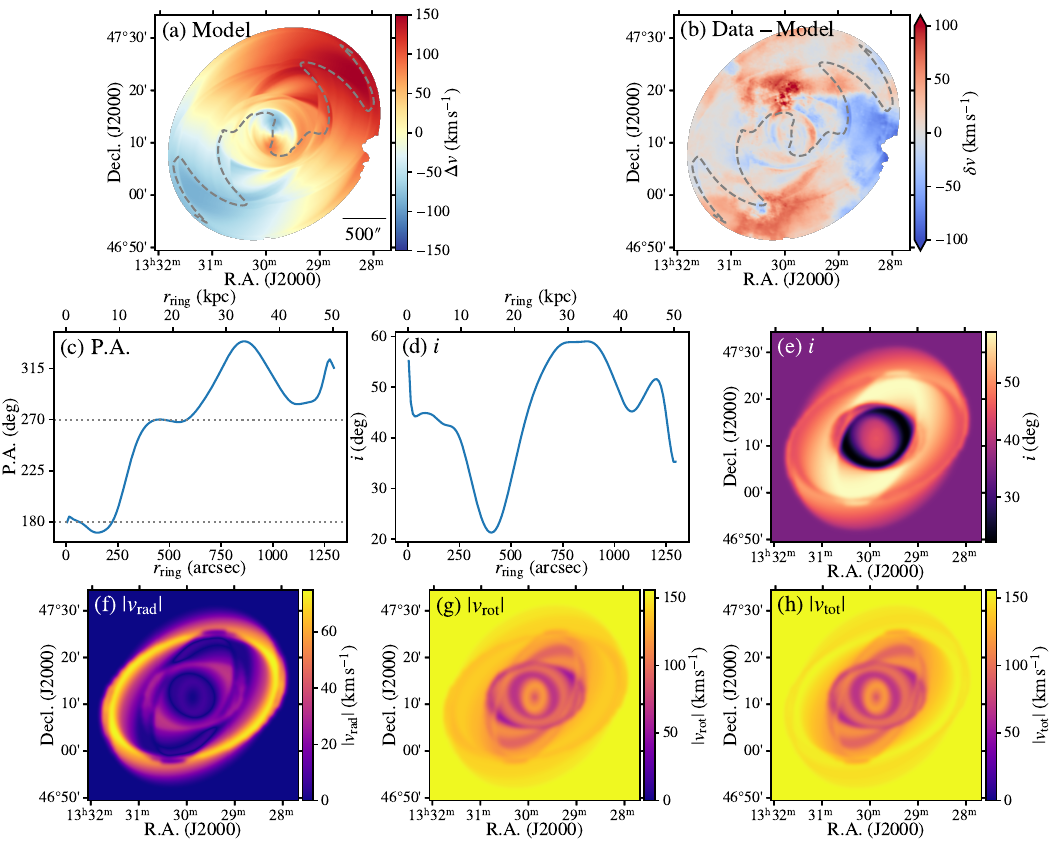}
  \caption{
    The full results of tilted-ring modeling using \emph{BBarolo}.
    (a)~The moment-1 map of the best-fitting model.
    The gray dashed line traces the major-axis ends of the rings.
    (b)~The residual between the data and model moment-1 map.
    (c)~The profile of the position angle of the receding side, starting from the north and increasing counterclockwise.
    (d)~The profile of inclination angle.
    (e)~The reconstructed inclination map.
    (f--h)~The reconstructed maps of radial (\vrad), rotational (\vrot), and total (\vtot) velocities.
  }
  \label{fig:ring_model_app}
\end{figure*}

Meanwhile, some characteristics of the real galaxy data may assist in increasing the actual resolution.
The ISM structures usually have a smooth transition between different scales and different densities, as indicated by the power-law power spectra \citep{2004ARA&A..42..211E}.
MEM requires a maximum entropy relative to the reference image, thus helping to give the smoothest solution that at the same time has the lowest residual rms levels.
The mock test in \autoref{app:ssec:recovery} reveals that effectively, a $2\sigma$ \NHi[lim] of \qtyNHilim is achieved at a high resolution of \qtyVLAfwhm, better than the S/N-based instrumental resolution.
A likely reason is that the intrinsic spatial scale for the \qtyNHilim \Hi gas is much larger than the instrumental resolution.
This is the same underlying principle for the method of linearly combining interferometric and single-dish data in the Fourier space \citetext{\citealp{2002ASPC..278..375S}; \citetalias{2024ApJ...973...15W}}:
because the large-scale \Hi missed by interferometers is so well-resolved by the single-dish telescope, an image at the single-dish resolution has no essential difference from the one at interferometric resolution.
Therefore, adding the missing \Hi back to the interferometric-only data results in a combined cube with the effective resolution of the interferometric clean beam.
Compared to the linear combination method, the joint deconvolution ensures a smoother transition in phases and scales between two types of data.

In summary, we estimate that at an \NHi level of \qtyNHilim, to reach a $2\sigma$ confidence for a single \los, the individual resolution element should be \qtyFWHMavg (\qtyphysFWHMavg[]).
The intrinsically large scale of low-\NHi gas helps to reach a higher resolution, e.g., \qtyVLAfwhm as used in the main part of this paper, as supported by our mock tests in \autoref{app:ssec:recovery} where the median values show good accuracy.
Other factors may also influence the resolution, such as the pointing and calibration uncertainties, which we do not consider here.

\subsection{Comparison to the Data Before Joint Deconvolution and to the Result of Linear Combination}
\label{app:ssec:mom0_comp}
The jointly deconvolved moment-0 map (\autoref{fig:mom0_comp}a) contains most of the large-scale structure detected by FAST (\autoref{fig:mom0_comp}c) with the resolution comparable to the interferometric THINGS data (\autoref{fig:mom0_comp}b).
Compared with the result of linearly combining the FEASTS data and the recleaned THINGS model using the procedure of \citetalias{2024ApJ...973...15W} (\autoref{fig:mom0_comp}d), the new joint deconvolution method provides a more natural transition from high- to low-\NHi regions and reveals more details at the outskirts.

\section{Deprojected Velocity and Inclination Maps}
\label{app:sec:vfieldmap}
The moment-1 map and the residual of the final model are shown in \autoref{fig:ring_model_app}(a) and~(b), and the profiles of \PA and $i$ are given in panels (c) and~(d).
Except for the northern region where M51b encounters M51a, the velocity field is rather well recovered.

By tilted-ring modeling, we aim to obtain the values of \Hi velocity without projection effects.
Such reconstructed velocity maps are shown in \autoref{fig:ring_model_app}(f--h).
For \vrot and \vrad, we first smoothed the fitted radial profile with a median filter, the kernel size being \ang{;;50} (five annuli).
Then, we sampled on each annulus with a separation of the annulus width (\ang{;;\sim10}), linearly interpolated the samples onto a grid, and finally smoothed the grid with a kernel of $\text{FWHM}=\ang{;;30}$.
By doing this, the values at the overlapping annuli are averaged out.
The total velocity is calculated as $\vtot=\sqrt{\vrot^2+\vrad^2}$.
A similar map of inclination $i$ is in panel~(e).

In the literature, velocity field fitting for M51a was typically limited within the relatively unperturbed part within \ang{;;\sim300} \citep[e.g.,][]{2014ApJ...784....4C,2014PASJ...66...77O,2019A&A...622A..58S,2023MNRAS.524.1560S}.
They generally reported a \PA of \ang{\sim170;;}, consistent with ours, but tended to estimate a lower $i$ of \ang{20;;}--\ang{35;;} in this radial range, possibly limited by the boundary at \ang{;;\sim300} where $i$ changes abruptly.
The maximum rotation velocity in their fittings typically reaches \qty{150}{\km\per\s}, while in our fitting the value is \qty{\sim125}{\km\per\s} within \ang{;;300}.
Our fitting is not contradictory to these previous results considering the different $i$ values.

\section{Spectral Component Fitting Procedures}
\label{app:sec:bigauss_decomp}
In the following, we provide the implementation of decomposing the diffuse- and dense-\Hi components mentioned in \autoref{ssec:bigauss_decomp}.
\begin{enumerate}
  \item The spectrum \Iv should have at least four channels brighter than \num{0.5} times the interferometric noise level of \sigint, and the peak intensity \Imax needs to be higher than $2\sigint$.
        Otherwise, this \los is considered to have only diffuse \Hi.
  \item A channel mask (mask~1) selecting those brighter than $1.5\sigint$ are generated.%
        \footnote{
          This channel mask and the $0.5\sigint$ one below, i.e., masks 1 and~2, have both had one-channel binary closing and dilation successively applied to them, so as to link and extend adjacent features.
        }
        If all the selected channels in mask~1 are connected, we decide there is only one dense Gaussian component in this \los, or else double-Gaussian fitting will be conducted.
        \begin{enumerate}
          \item For single-Gaussian fitting, the moment 0, 1, and 2 values are used to generate the initial guess.
          \item For double-Gaussian fitting, the two regions with highest intensities are used to determine the initial guess for each of the two Gaussian models, respectively.
        \end{enumerate}
  \item We use \emph{SciPy} \texttt{optimize.curve\_f{}it} \citep{2007WRR....43.3423V} to get the least-squares optimized models for the region within mask~2 (brighter than $0.5\sigint$).
        The model function is one or two positive Gaussian function(s) of dense \Hi, plus a positive constant baseline.
        The data cubes for the dense and diffuse components are generated accordingly.
\end{enumerate}
All of the fittings converge within \num{170} iterations.
On the whole, the \los line structures are relatively simple.
Among the \los sample, only \qty{11}{\percent} of \los{}s require a decomposition, \qty{7}{\percent} of which (\qty{0.8}{\percent} of the \los sample) contain two Gaussian components.

\section{Averaged Column Density Profile of the FEASTS Sample}
\label{app:sec:FEASTS_NHi}
\begin{figure}
  \centering
  \includegraphics{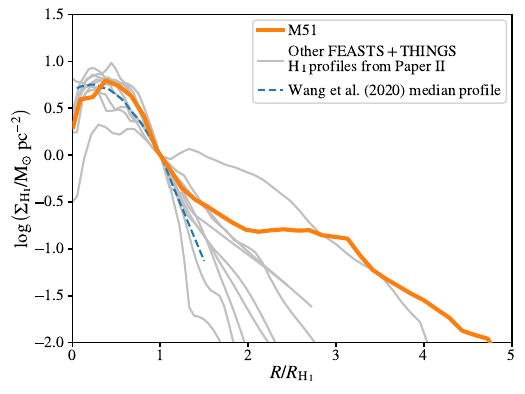}
  \caption{
    Comparing the \Hi surface density (\SigHi) profiles of M51 (orange thick line) and nine other galaxies (gray lines), normalized by \Hi radius \RHi of \qty{1}{\Msun\per\pc\squared}.
    These galaxies have all been observed by both THINGS and FEASTS, and the profiles are taken from \citetalias{2024ApJ...973...15W}, where THINGS and FEASTS data are linearly combined.
    The profile of M51 is almost flat between $\numlist{1.5;2.5}\RHi$ with $\SigHi\approx\qty{0.3}{\Msun\per\pc\squared}$, different from most of the other galaxies.
    The median \Hi profile from compiled interferometric observations are given as the blue dashed line for reference \citep{2020ApJ...890...63W}.
  }
  \label{fig:SigHI_prof}
\end{figure}

We compare the \Hi surface density (\SigHi) profiles of M51 and nine other galaxies in \autoref{fig:SigHI_prof} (data from \citetalias{2024ApJ...973...15W}).
M51 has a distinct almost flat profile between $\numlist{2;3}\RHi$ at $\SigHi\approx\qty{0.2}{\Msun\per\pc\squared}$, and the slope at larger radii is also shallower than other profiles.
Here, the \Hi radius \RHi is defined at an \Hi surface density of \qty{1}{\Msun\per\pc\squared}.

\section{Channel Maps of Deconvolved Model and Residuals}
\label{app:sec:chan_map}
\begin{figure*}
  \centering
  \includegraphics{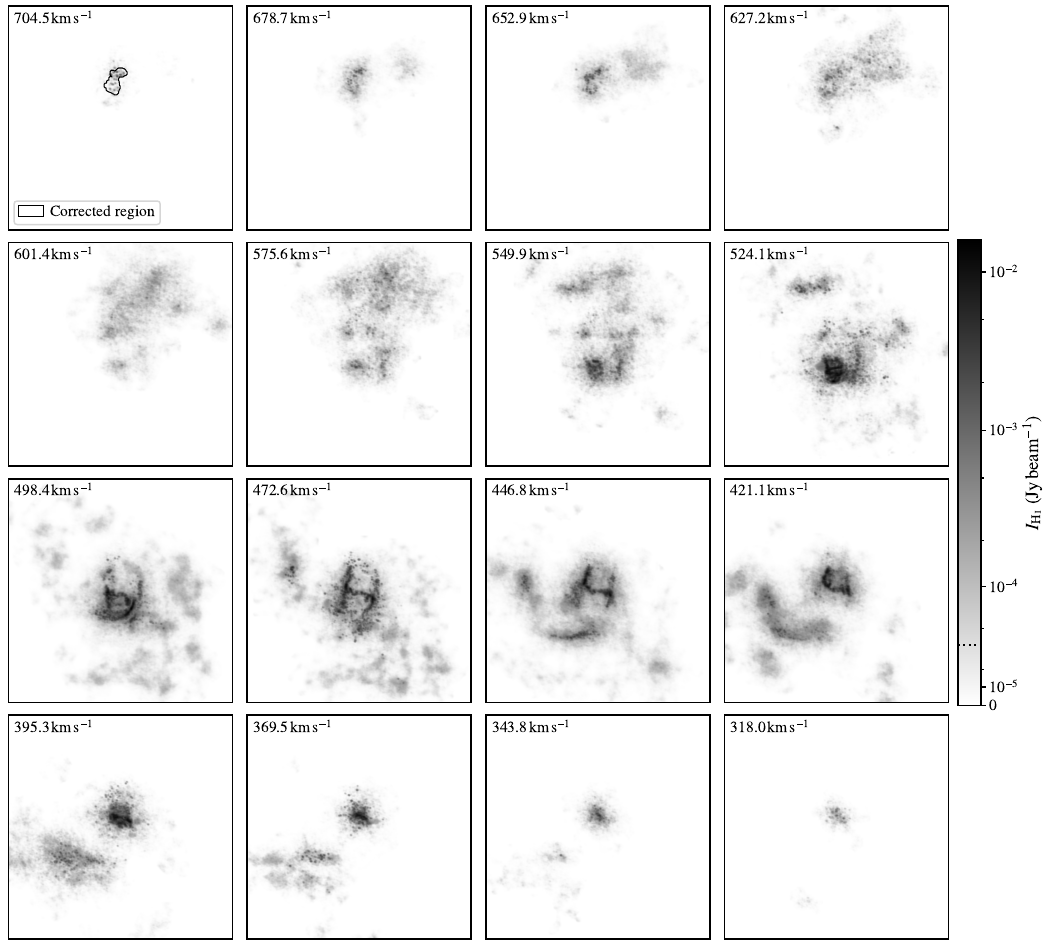}
  \caption{
    The channel maps of the jointly deconvolved cube of M51, combining the FEASTS and THINGS data.
    Every fifth channel is shown (channel width: \qtychanwidth).
    The color coding is in asinh scale, with the luminosity below $\sqrt{\sigint\sigsd}=\qty{3.665e-2}{\mJy\per\bvla}$ (the horizontal black dotted line on the color bar) being approximately linear.
    In the first channel map, the black contour indicates the region where we correct the THINGS visibility using \citet{1990AJ....100..387R} data (see \autoref{ssec:vla_data}).
  }
  \label{fig:chan_cube}
\end{figure*}

\begin{figure*}
  \centering
  \includegraphics{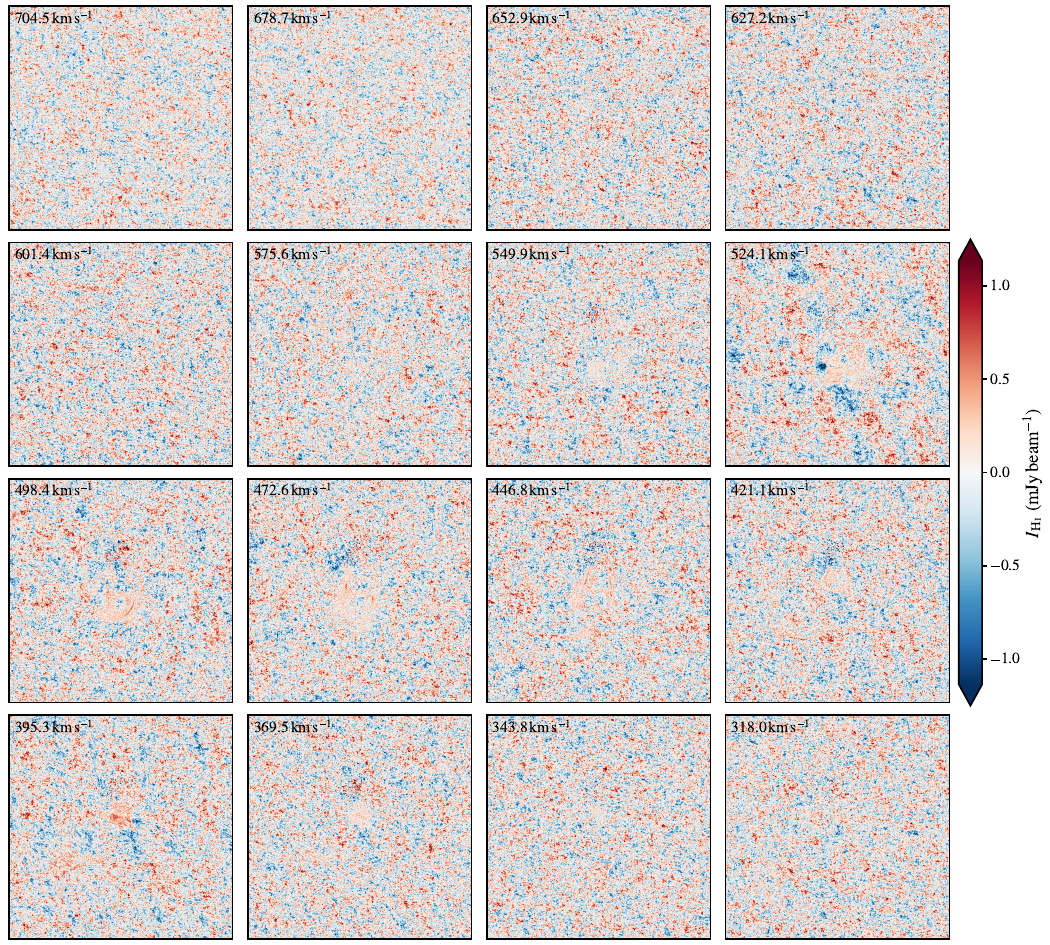}
  \caption{
    The same as \autoref{fig:chan_cube}, but for the interferometric residual (with PB attenuation).
    The color coding is linear, with $3\sigint=3\times\qty{0.379}{\mJy\per\bvla}$ as the limit.
  }
  \label{fig:chan_intres}
\end{figure*}

\begin{figure*}
  \centering
  \includegraphics{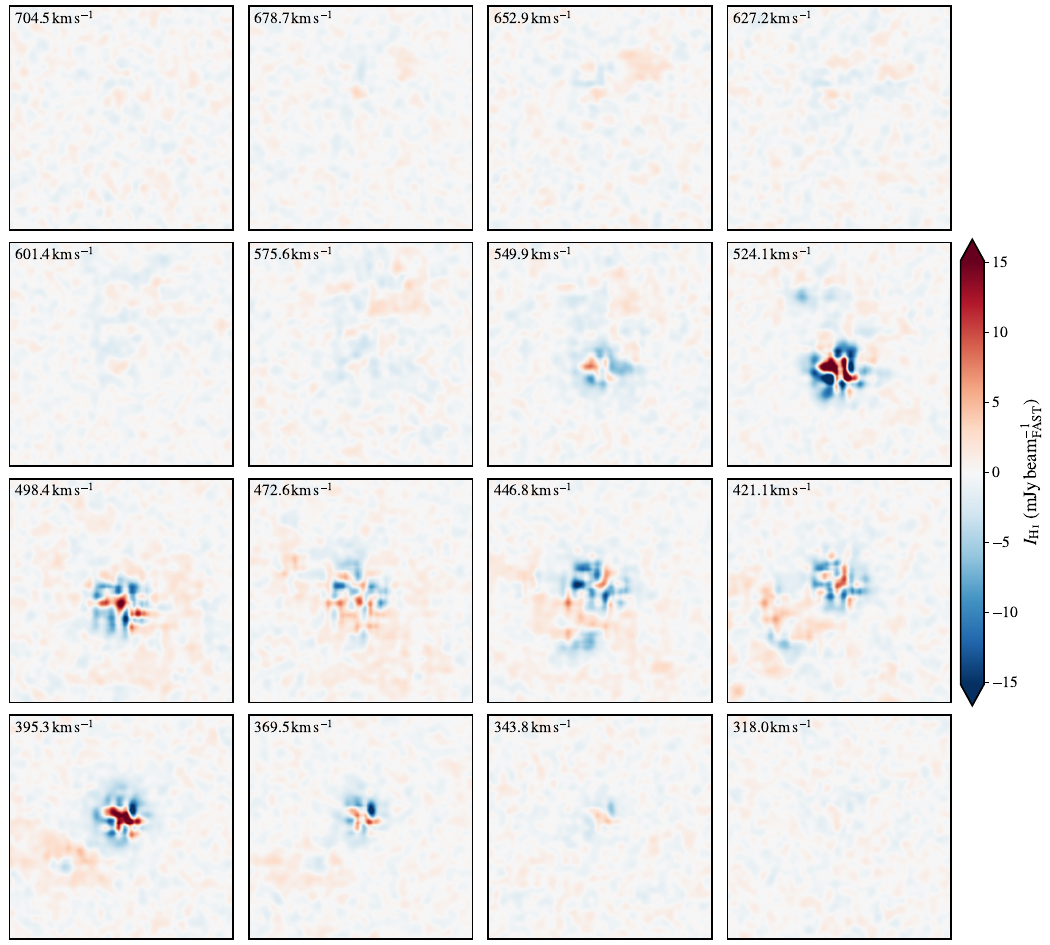}
  \caption{
    The same as \autoref{fig:chan_cube}, but for the single-dish residual.
    The color coding is linear, with $3\sqrt{\sigint\sigsd}=3\times\qty{5.05}{\mJy\per\bfast}$ as the limit.
  }
  \label{fig:chan_sdres}
\end{figure*}

In Figures \ref{fig:chan_cube}--\ref{fig:chan_sdres}, we provide the every fifth channel map of the jointly deconvolved model, interferometric residual, and single-dish residual cubes.

\bibliography{ms}

\end{document}